\documentclass[final,3p,times,twocolumn]{elsarticle} 

\usepackage{amssymb}
\usepackage{amsmath}

\usepackage{float} 

\usepackage{graphicx} 

\usepackage{booktabs} 

\usepackage{multirow} 

\usepackage[hyphens]{url}
\usepackage{microtype}

\begin{document}

\begin{frontmatter}



\title{Anthropomorphism on Risk Perception: The Role of Trust and Domain Knowledge in Decision-Support AI} 


\affiliation[cuhksz]{organization={School of Management and Economics, The Chinese University of Hong Kong, Shenzhen},
                     city={Shenzhen},
                     country={China}}

\affiliation[hkustgz]{organization={Computational Media and Arts, The Hong Kong University of Science and Technology (Guangzhou)},
                      city={Guangzhou},
                      country={China}}

\author[cuhksz]{Manuele Reani\corref{cor1}}
\ead{reanimanuele@cuhk.edu.cn}

\author[hkustgz]{Xiangyang He}

\author[cuhksz]{Zuolan Bao}

\cortext[cor1]{Corresponding author}



\begin{abstract}
Anthropomorphic design is routinely used to make conversational agents more approachable and engaging. Yet its influence on users’ perceptions remains poorly understood. Drawing on psychological theories, we propose that anthropomorphism influences risk perception via two complementary forms of trust, and that domain knowledge moderates these relationships. To test our model, we conducted a large-scale online experiment (N = 1,256) on a financial decision‑support system implementing different anthropomorphic designs. We found that anthropomorphism indirectly reduces risk perception by increasing both cognitive and affective trust. Domain knowledge moderates these paths: participants with low financial knowledge experience a negative indirect effect of perceived anthropomorphism on risk perception via cognitive trust, whereas those with high financial knowledge exhibit a positive direct and indirect effect. We discuss theoretical contributions to human‑AI interaction and design implications for calibrating trust in anthropomorphic decision‑support systems for responsible AI.
\end{abstract}




\end{frontmatter}




\section{Introduction} \label{inro}

Digital assistants powered by large language models (LLMs) are increasingly mediating decisions in customer service, education, healthcare, and finance. To make these systems feel more social and relatable, designers frequently employ anthropomorphic cues -- e.g., human‑like faces, names, avatars, and specific styles of verbal and non-verbal communication (typical of human-to-human interactions). Anthropomorphic design in Artificial Intelligence (AI) seems to elicit greater engagement, adoption, trust, arousal, and likability~\citep{bickmore2005social, song2020trust}. Anthropomorphic cues, thus, greatly affect human-AI interaction and represent a powerful lever that AI designers can exploit to influence users.

However, humanising AI can create unrealistic expectations, ethical concerns, and unwanted consequences. For instance, anthropomorphic appearance seduces users into attributing more agency and competence to AI than warranted~\citep{peter2025benefits}. In fact, some evidence suggests that anthropomorphism tends to increase trust and reduce risk perception~\citep{cohn2024believing}, a concerning effect that can be problematic, especially in high-risk contexts (e.g., financial investment and medical decision-making). 

Empirical evidence shows that, in certain circumstances, users can follow AI advice that contradicts other available information, or even their own judgment~\citep{klingbeil2024trust}. Recent work found that participants consistently regard AI-generated content, despite being correct or incorrect, as more trustworthy, valid, and satisfying than advice written by doctors~\citep{shekar2025people}. In high-stakes domains such as healthcare, placing excessive trust in online information (from search engines, social media, or AI) can lead users to defer to recommendations even when they are wrong, resulting in worse outcomes than if they had ignored the information~\citep{bbc2016baidu}. These results further highlight the importance of understanding risk perception in the context of AI-aided decision-making.

Responsible AI design, therefore, requires a nuanced understanding of how anthropomorphism can affect trust and the perception of risk. Prior studies in the field of human-AI collaboration have examined anthropomorphism, trust, and risk perception separately, but few integrate these constructs into a comprehensive model, especially in the context of AI~\citep{Aubel2022}. To complicate the puzzle further, domain expertise has been shown to affect how people perceive risk. In some research, people with greater domain knowledge perceive lower risk because their deeper understanding reduces reliance on heuristic cues~\citep{brauner2024misalignments}. Other studies, however, report that, in high-risk settings, risk perception increased with higher expertise~\citep{said2025artificial}. This suggests that domain knowledge may moderate the effects of anthropomorphic design on trust and risk perception in different ways. To address these gaps, and shed light on the interplay between these factors, we propose and test a moderated‑mediation model linking anthropomorphic design, trust, and risk perception. 

The question of how a more anthropomorphic style might be beneficial/detrimental in scenarios where rationality and a lack of emotion are preferred (e.g., finance) is relevant and timely, given the rise of chatbots and LLMs. The relationships between anthropomorphism, trust, and risk perception have been well-documented, although not in the same study -- see Section~\ref{relwork}. Moreover, including domain knowledge in anthropomorphism research is intuitive and novel.

Our contributions are fourfold: (1) we probe into the relationship between anthropomorphism design and risk perception by distinguishing cognitive and affective trust as parallel mediators of the anthropomorphism-risk relationship; (2) we examine domain knowledge as a moderator; and (3) we provide empirical evidence using a large‑scale experiment with rigorous analyses, including multi‑group SEM and robustness checks. Finally, (4) we discuss the design implications of such a human-AI interaction model. 

\section{Related Work}\label{relwork}
\subsection{Benefits of Anthropomorphic Design}
Anthropomorphism refers to the human tendency to attribute human qualities to non‑human entities~\citep{Epley2007, Airenti2015, zlotowskiAnthropomorphismOpportunitiesChallenges2015a}. Besides people's natural tendency to anthropomorphize technology, this tendency can be amplified by the scientific community's effort of "designing anthropomorphism"~\citep{Salles2020}. Thus, anthropomorphic design is the science and practice of making a technological artefact more human-like or making such an artefact behave like a human. Research has shown that the anthropomorphic design of AI has several benefits, including an increased adoption of and intention to use such technologies~\citep{Bruce2002, Stroessner2019}. Designers leverage anthropomorphic cues – i.e., names, avatars, and language – to evoke social presence and engagement~\citep{Seeger2021, Bao2022, klein2023impact}. Such a design can positively affect business and bring a better experience to users. 

However, the same techniques that make AI systems more engaging can also have unintended side effects. By triggering social heuristics and “computers as social actors” responses~\citep{Nass1994}, anthropomorphic cues can lead people to ascribe agency, benevolence, and competence to AI systems that exceed their actual capabilities~\citep{waytz2014mind,reani5222775fundamental,hasan2025dark}. In low‑stakes settings, this may primarily influence likability, but in high‑stakes decision‑support contexts such as healthcare and finance, these shifts in perception may systematically bias how people judge the trustworthiness and riskiness of AI advice. This tension between user experience benefits and potential miscalibration of trust and risk perception motivates a closer examination of anthropomorphic design in decision‑support AI.
\subsection{Cognitive and Affective Trust in AI}
Trust is a multifaceted construct. Specifically, two dimensions of trust have been identified. Cognitive trust involves rational evaluations of an agent’s ability, reliability, and transparency; affective trust arises from warmth, empathy, and emotional bonds~\citep{Johnson2005,li2024developing}. 

Research on social robots and AI agents shows that anthropomorphism can affect trust. For instance, some research shows that an anthropomorphic appearance (human-like voice and faces) increases perceived trust and makes AI systems seem more competent and warm; with warmth signaling benevolent intent, while competence signaling (expected) ability~\citep{li2024developing}. The reason why this happens is not fully understood. Recent work emphasizes the role of anthropomorphism in improving trust via social presence and provides an explanation, albeit partial, by focusing on affective trust~\citep{li2023influence}. Cognitive trust, which is linked to competence and expectations, is not accounted for by this explanation. Some earlier research has already shown that users interacting with anthropomorphized digital assistants (e.g., human names and communication style) expect them to behave like competent human aides~\citep{luger2016like}. Moreover, in other domains, research has shown that (cognitive) trust mitigates uncertainty and, thus, reduces perceived risk~\citep{zhou2025mediation}. Thus, fragmented research has overall shown the benefit of AI anthropomorphism on both cognitive and affective trust, but not how these two dimensions of trust are affected differently by such designs.    

\subsection{Risk Perception in Anthropomorphic Design}
According to the so-called psychometric paradigm of risk perception pioneered by Paul Slovic, risk can be defined as the "quantitative measure of hazard consequences [i.e., threats to humans and what they value] expressed as conditional probabilities of experiencing harm"~\citep{Slovic1985}. The term "risk perception" can be simply understood as people's intuitive judgment, measurement, and evaluation of hazardous activities -- e.g., financial investments -- and the interaction with new technologies -- e.g., AI systems~\citep{Slovic2005}. In our study, risk perception refers to the user's judgment of how risky it would be to follow an AI advisor's recommendation (e.g., financial investment), which could carry serious negative consequences (e.g., financial loss).
   
Early work in non-AI contexts already anticipated that anthropomorphic design (placing human eyes on machines used for gambling) can affect risk perception \citep{Kim2011}. Such work demonstrated that anthropomorphism increases risk perception for people with low power and decreases it for those with high power -- here ``power'' is a situational psychological state indicating how much control and agency a person feels they have at a given moment.~\citet{Morana2020} studied the effect of different levels of chatbot anthropomorphism (human names, avatars, and the manipulation of language cues) on the likelihood of following the recommendations given by the robo-advisor in an investment consultation scenario. In their research, risk perception was not the primary focus of the study; rather, it was implicitly linked to the likelihood of following the recommended financial advice. Nevertheless, they did measure individual risk level as a control factor. Thus, they already anticipated (without directly testing) that the manipulation of anthropomorphism in chatbot systems may indeed affect risk perception~\citep{Morana2020}.~\citet{Aubel2022} examined how visual anthropomorphism influences risk perception in a banking context. They manipulated only the agent's appearance (from more machine-like to more human-like) and then assessed how risky users perceived the interaction to be. The results show that anthropomorphism, indeed, affects risk perception in a non-monotonic (uncanny-valley) pattern. Finally,~\citet{cohn2024believing} studied how users anthropomorphize LLM systems (through the manipulation of linguistic/modality cues) and how they trust the information the system provides. The results of this study show that perceived anthropomorphism indeed increases the perception of accuracy of the information (thus, related to competence, which is an element of cognitive trust) and, in turn, reduces risk ratings over the consequences of potential misinformation. These results show a clear effect of anthropomorphism on risk perception. Nevertheless, this study did not take into account domain knowledge, which appears to play a moderating role in risk contexts. Indeed, in recent research, it was found that expertise increases risk perception~\citep{said2025artificial} and that experts rely less on anthropomorphic cues~\citep{brauner2024misalignments}.

\subsection{Domain Knowledge in High-stakes Domains}
Finance is a paradigmatic high‑stakes domain where misaligned trust and risk perception have direct monetary and welfare consequences. Commercial robo‑advisors and banking chatbots are increasingly deployed with human‑like names, avatars, and conversational styles (e.g., “Erica”, “Cora”) to emulate a human financial advisor and lower adoption barriers~\citep{hyunbaekAiRoboadvisorAnthropomorphism2023,schreibelmayr2023first}. Empirical work suggests that such cues can meaningfully change how people evaluate financial advice: anthropomorphic robo‑advisors can increase perceived usefulness, trust, and willingness to invest, partly by raising certainty about the advisor’s competence and lowering perceived risk~\citep{hyunbaekAiRoboadvisorAnthropomorphism2023,hildebrand2021conversational}. Recent experimental evidence further indicates that anthropomorphic framing of a financial AI advisor can reduce perceived risk via increases in both cognitive and affective trust, especially among lay investors~\citep{reani5222775fundamental}. Together with findings that experts are more skeptical of anthropomorphic AI and perceive higher risk in AI‑supported decisions~\citep{said2025artificial,brauner2024misalignments}, this body of work suggests that anthropomorphic design may systematically benefit novice users while simultaneously creating conditions for overtrust and miscalibrated risk perception in financial decision‑support.

To conclude, a growing but still fragmented body of work across chatbots, social robots, and financial decision‑support systems suggests that anthropomorphic cues may increase trust and, in some populations, reduce perceived risk. However, existing studies typically focus either on average effects of anthropomorphism on trust~\citep{Morana2020,Seeger2021,Waytz2010,hyunbaekAiRoboadvisorAnthropomorphism2023} or on its impact on risk perception in specific scenarios~\citep{Kim2011,Aubel2022,cohn2024believing}, and rarely model the underlying trust processes or their interaction with user expertise. In particular, prior research has not systematically examined how cognitive and affective trust jointly mediate the link between perceived anthropomorphism and risk perception, nor how domain experts and lay people may respond differently to the same anthropomorphic cues in high‑stakes financial advice. Our study addresses this gap.

\subsection{Research Gaps and Objective}
A human-like presentation can be counterproductive in settings where analytical detachment is valued, such as finance, making this question especially salient in the era of LLMs and chatbots. Prior work has linked anthropomorphism to trust~\citep{Morana2020} or perceived risk~\citep{Aubel2022}, though rarely within a single integrated study. Adding domain knowledge as a moderating factor is both intuitive and, to our knowledge, relatively novel in this line of research. Current literature lacks an integrative human-AI interaction model that connects anthropomorphic design, dual trust processes, and risk perception while accounting for user expertise. Most prior studies in anthropomorphic AI examine either cognitive or affective routes, seldom both simultaneously. Few investigate how domain knowledge moderates these routes or test models using large samples. Moreover, many studies focus on embodied robots or low‑stakes conversational agents; little is known about anthropomorphic decision‑support systems in high‑stakes domains (e.g., financial investment). Our work addresses these gaps by proposing a moderated‑mediation model and empirically testing it using a large sample while manipulating anthropomorphic design in an AI system. 

In this work, we consider the effect of perceived domain knowledge on the relationship between AI anthropomorphism and human perception (risk perception and trust), for a holistic understanding of how humans react to humanized bias.

\section{Theory and Hypotheses} \label{theory} 
The effect of the anthropomorphic design of AI agents on perceived anthropomorphism (the human perception of non-human entities -- see~\citet{Bartneck2009}) is well documented~\citep{Bao2022, sun2024anthropomorphism}. Thus, we consider such a design an antecedent of perceived anthropomorphism, which is the actual predictor of our framework. As suggested in previous work~\citep{Kim2011, adam2019investment, Morana2020, Aubel2022}, the design of technological artifacts, including AI, especially the use of cues that make such technologies look or behave like humans, affects risk perception. This, however, appears to be contingent on high-risk environments~\citep{adam2019investment}. Thus, there are reasons to believe that perceived anthropomorphism affects risk perception in high-risk scenarios (e.g., financial investing), albeit it is not clear whether this effect is direct or mediated by trust. 
\citet{Kim2011}’s results are mixed since they found anthropomorphism to increase risk perception for some individuals and reduce it for others (see the model development in the next sub-section). Likewise,~\citet{Aubel2022} reported that high anthropomorphism increases risk perception, but this relationship is dependent on the design of AI. Therefore, our initial hypothesis, the base hypothesis, is confirmatory and serves the purpose of assessing whether there is a direct effect, either positive or negative, of perceived anthropomorphism on risk perception, thus:

\begin{itemize}
    \item H1: Perceived anthropomorphism of chatbots will affect risk perception (non-directional).
\end{itemize}

It is worth noting that, in our investigation, we focus on perceived anthropomorphism, rather than its design, for three reasons. First, trust and risk perception are psychological responses to how human-like AI feels to a user. Two participants can see the identical interface (i.e., same design cues), yet experience different human-likeness~\citep{seeger2021designing,Bao2022}. Rather than the design code per se, it is that experienced human-likeness (i.e., perceived anthropomorphism) that affects other perceptions (e.g., trust and risk perception). Second, traits such as individual anthropomorphism tendencies or prior chatbot experience may modulate how strongly the same cues translate into perceived human-likeness. Perceived anthropomorphism naturally absorbs this heterogeneity; anthropomorphic design cannot. Finally, perceived anthropomorphism travels better than a specific cue recipe. Practitioners rarely reproduce the exact design combinations used in experiments; what generalizes is the level of perceived human-likeness. Modeling perception rather than design makes findings portable to future systems that achieve similar perceived anthropomorphism with different cue mixes.

\subsection{Research Framework}
Studies investigating the relationships between anthropomorphism and trust have shown that users are more likely to trust anthropomorphized agents to make decisions -- e.g., investing~\citep{Waytz2010}. \citet{Morana2020} found that an increased level of anthropomorphism of chatbots increases users' trusting beliefs towards those chatbots. Beyond these individual studies, a broader body of research in HCI, social robotics, and decision-support systems has consistently shown that anthropomorphic cues (e.g., human-like names, faces, and conversational styles) strengthen users' perceptions of competence, benevolence, and integrity and thereby increase trust in AI agents~\citep{Nass2000,qiu2009evaluating,pak2012decision,droeslerevisserAlmostHumanAnthropomorphism2016a}. Meta-analytic evidence in human–robot interaction further indicates a robust positive effect of anthropomorphism on user trust across a wide range of artificial agents~\citep{roesler2021meta}. In general, these findings support the view that anthropomorphism in AI tends to increase user trust~\citep{Seeger2021}.

In addition, earlier research has extensively shown that trust is negatively correlated with risk perception, especially those elements of trust related to competence~\citep{Siegrist2005}. If anthropomorphism increases trust and trust reduces risk perception, then, by transitivity, anthropomorphism should reduce risk perception.   

\subsubsection{A Dual Route to Trust.} 
Previous research on trust separates cognitive trust from affective trust~\citep{Johnson2005}. On the one hand, cognitive trust is knowledge-driven and includes elements of the trustee’s competence, expertise, reliability, and perceived performance~\citep{moorman1992relationships, rempel1985trust}. On the other hand, affective trust is related to emotional connections and the trustee’s benevolence and is the confidence placed in another person on the basis of feelings generated by the level of care and concern the person demonstrates~\citep{johnson1982measurement, rempel1985trust}. In our analytical model of risk-relevant perceptions (the research framework), trust is seen as an intermediate psychological state (a mediator) between perceived anthropomorphism and risk perception. 

\subsubsection{Cognitive Trust as a mediator.} 
The manipulation of anthropomorphism, especially the utilization of certain communication cues – e.g., the use of informal expressions in communication~\citep{Seeger2021} – can make the advisor look friendly but, at the same time, less competent or less professional~\citep{boutet2021emojis, glikson2018dark}. In certain contexts (e.g., finance), a machine-like style of communication may look more professional (albeit less warm and emphatic) and, thus, may convey more competence than an informal and friendly style of communication~\citep{kim2012anthropomorphism,munnukka2022anthropomorphism,toure2015or}. It follows that increasing anthropomorphism of advisory AI with specific design choices may reduce perceived competence. In turn, low competence has been shown to increase risk perception~\citep{veres2009competence}. 

In the trust model described earlier, competence is captured by the cognitive component of trust~\citep{Johnson2005}. Thus, manipulating anthropomorphism in AI systems that are used to communicate information that bears risk to users may affect cognitive trust to some extent. In our model, we posit that cognitive trust is a mediator in the relationship between anthropomorphism and risk perception. Thus, we propose the following hypotheses:  

\begin{itemize}
    \item H2: Cognitive trust will mediate the effect that anthropomorphism has on risk perception.
    \begin{itemize}
        \item H2A: Anthropomorphism will increase cognitive trust.
        \item H2B: Cognitive trust will decrease risk perception.
    \end{itemize}
\end{itemize}

\subsubsection{Affective Trust as a mediator.}
In their research, Aubel\citep{Aubel2022}, speculated that the effect of anthropomorphism on risk perception might be due to an “affect heuristic” for risk perception -- i.e., an emotional response to the stimuli -- a concept that is closely related to the affective dimension of trust. In the case of anthropomorphic AI, a human-like communication style better resonates with our affective state, increasing our trust in the source of the message~\citep{Seeger2021}, provided that such a source is competent and professional. It follows that perceived anthropomorphism may positively affect affective trust. However, whether this, in turn, lowers risk perception is not clear from the literature. For this dimension, we posit the following three hypotheses:

\begin{itemize}
    \item H3: Affective trust will mediate the effect that anthropomorphism has on risk perception.
    \begin{itemize}
        \item H3A: Anthropomorphism will increase affective trust.
        \item H3B: Affective trust will affect risk perception (non-directional).
    \end{itemize}
\end{itemize}

\subsubsection{Contradiction in the literature and the role of Domain Knowledge.} 
As reported above, by the transitivity principle, if anthropomorphism increases trust and trust decreases risk perception, then anthropomorphism should reduce risk perception. However, the research by~\citet{Kim2011} found that, although anthropomorphized objects generally generate a positive attitude towards those objects, when they bear risk, the effect may not be positive. This suggests that, in high-risk settings, artifacts resembling humans can increase, for certain individuals, the risk perception of what those artifacts are prompting the user to do (e.g., gambling). Similarly,~\citet{Aubel2022} found a greater risk perception when anthropomorphism is high. However, it is not clear why anthropomorphism increases trust~\citep{Morana2020, Seeger2021, Waytz2010} and also risk perception~\citep{Aubel2022, Kim2011}, given that trust and risk perception tend to be negatively correlated~\citep{johnson1995presenting}. We posit that domain knowledge is the potential candidate that may shed light on these relationships, given that, in high-risk settings, higher expertise increases risk perception~\citep{said2025artificial} and negatively affects trust in AI-based decision support systems~\citep{bayer2022role}. Thus:

\begin{itemize}
    \item H4: Domain knowledge will increase risk perception.
    \begin{itemize}
        \item H4a: Domain knowledge moderates the effect that anthropomorphism has on cognitive trust.
        \item H4b: Domain knowledge moderates the effect that anthropomorphism has on affective trust.
    \end{itemize}
\end{itemize}

\begin{figure*}[t]
    \centering
    \includegraphics[width=0.7\linewidth]{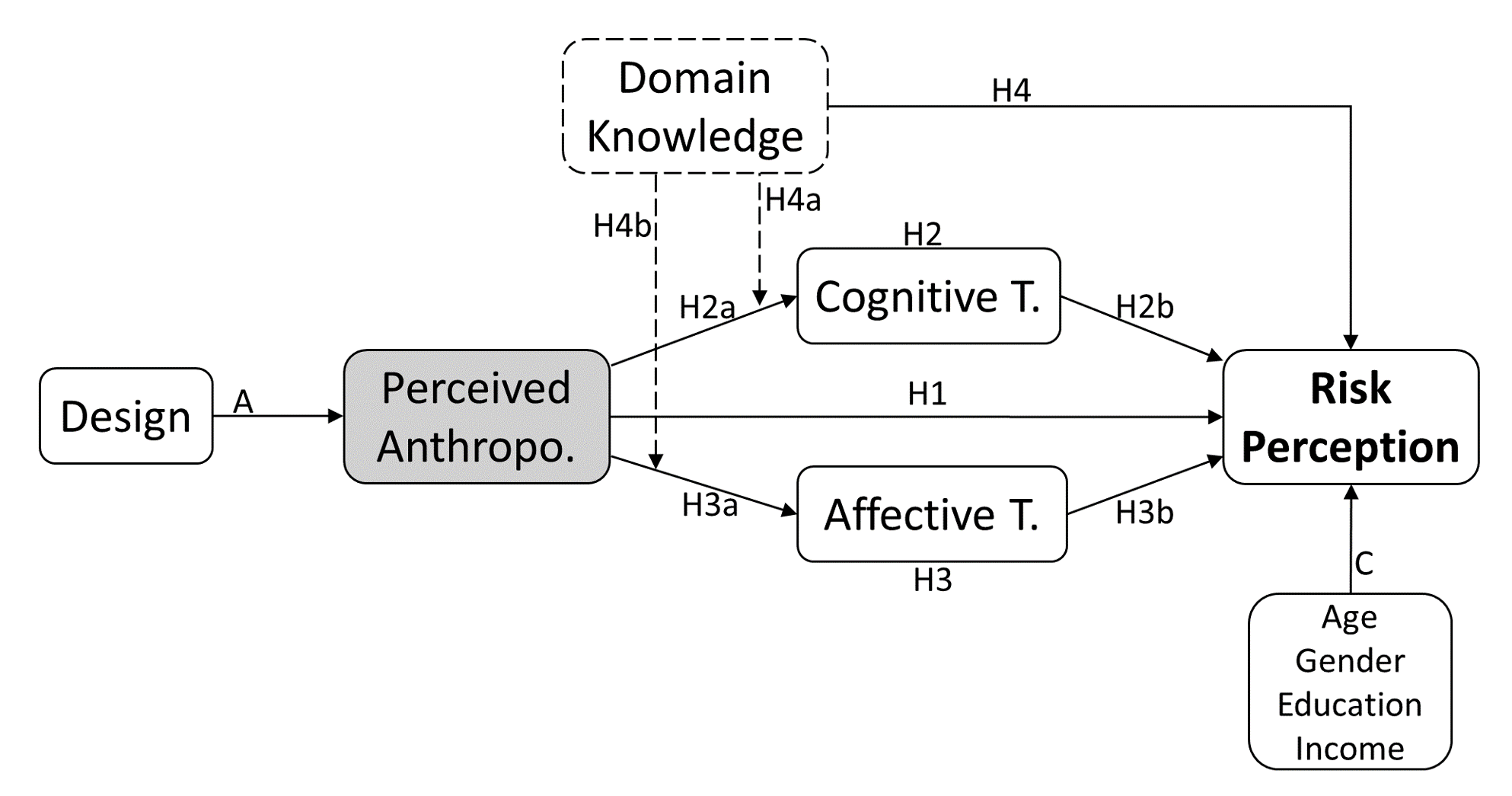}
    \caption{A path diagram showing relationships between variables Anthropomorphic Design, Perceived Anthropomorphism, Cognitive Trust, Affective Trust, and Risk Perception, with moderating variables Domain Knowledge and demographic factors (Age,
  Gender, Education, Income). Arrows labeled H1-H4 with sub-labels (H2a, H2b, H3a, H3b, H4a, H4b) show hypothesized relationships between the constructs.}
    \label{fig:moderated}
  \end{figure*}

These relationships and the complete research framework are shown in Figure~\ref{fig:moderated}, and the collection of hypotheses is shown in Table~\ref{tab:hypotheses}.

\begin{table*}[t]
  \centering
  \caption{Summary of hypotheses.}
  \label{tab:hypotheses}
  \resizebox{\textwidth}{!}{%
  \begin{tabular}{ll}
    \toprule
    \textbf{Hypothesis} & \textbf{Description} \\
    \midrule
    H1 & Perceived Anthropomorphism of chatbots will affect Risk Perception (non-directional). \\
    H2 & Cognitive Trust will mediate the effect that anthropomorphism has on Risk Perception. \\
    H2A & Anthropomorphism will increase Cognitive Trust. \\
    H2B & Cognitive Trust will decrease Risk Perception. \\
    H3 & Affective Trust will mediate the effect that anthropomorphism has on Risk Perception. \\
    H3A & Anthropomorphism will increase Affective Trust. \\
    H3B & Affective Trust will affect Risk Perception (non-directional). \\
    H4 & Domain Knowledge will increase Risk Perception. \\
    H4a & Domain Knowledge moderates the effect that anthropomorphism has on Cognitive Trust. \\
    H4b & Domain Knowledge moderates the effect that anthropomorphism has on Affective Trust. \\
    \bottomrule
  \end{tabular}
  }
\end{table*}

\subsection{Variable Definitions} \label{variables} 
To clarify the framework, here we describe our variables, what they represent, and how they are measured, including the reference for the questionnaires used.

\textbf{Anthropomorphic Design (AD)}: An 8-level design borrowed from~\citet{Bao2022} in which anthropomorphism was manipulated over 3 different cues (name, avatar, and communication style).

\textbf{Perceived Anthropomorphism (PA)}: This is our main predictor. For this measure, we used the popular Godspeed questionnaire developed by~\citet{Bartneck2009} on a 4-item 7-point Likert scale. Thus, anthropomorphism is the perception of how human-like the system feels and looks.  

\textbf{Cognitive Trust (CT)}: A five‑item sub-scale capturing the cognitive aspect of trust – see~\citet{Johnson2005}. This is defined as the cognitive route to trust related to the perceived credibility and competence of the AI system. 

\textbf{Affective Trust (AT)}: A five‑item sub-scale capturing the affective aspect of trust – see~\citet{Johnson2005}. This is defined as the affective (emotional) route to trust related to the perceived warmth and goodwill of the AI system.

\textbf{Domain Knowledge (DK)}: Participants rated their perceived knowledge of financial markets and their familiarity with financial investments on a 7‑point Likert item -- i.e., self-reported domain knowledge~\citep{chen2022informal}. The use of self-reported domain knowledge is not uncommon in behavioral research -- for a full discussion see~\citet{flynn1999short,brucks1985effects,chen2022informal}. Our model targets perception-driven outcomes -- i.e., perceived anthropomorphism, trust, and risk perception; Thus, a self-assessed (perceived) level of domain knowledge is the construct that best aligns with this paradigm.

\textbf{Risk Perception (RP)}: Assessed using a 7-point Likert item capturing the perceived risk of the recommended investment. Thus, RP is a subjective judgment of the potential negative consequences of relying on AI advice.~\citet{Holzmeister2020} used the same scale but in a different scenario. In our study, risk perception is the participant’s subjective judgment of how risky it would be to follow the investment recommendation provided by the AI advisor. Implicitly, this action could lead to severe negative consequences (e.g., money loss). Our focus was on the action the system recommends, not on how the presentation of statistical information affects this perception, as found in~\citet{Holzmeister2020}. Thus, the recommendation was presented in-text by the AI system, along with the justification for such a recommendation. 

We also measured demographics such as Age, Gender, Education, and Income as control variables. For the variables' description and corresponding references, see Table~\ref{tab:variables}. 

\begin{table*}[ht]
\caption{Variables descriptions, references, and uses in questionnaire.}
\label{tab:variables}
\resizebox{\textwidth}{!}{%
\begin{tabular}{llll}
\toprule
\textbf{Variable Name} & \textbf{Questionnaire name} & \textbf{Reference} & \textbf{Use} \\
\midrule
Anthropomorphic Design & (Exp. manipulation) & \citet{Bao2022} & Antecedent \\
Perceived Anthropomorphism & Godspeed & \citet{Bartneck2009} & Predictor \\
Cognitive Trust & CATQ & \citet{Johnson2005} & Mediator \\
Affective Trust & CATQ & \citet{Johnson2005} & Mediator \\
Domain Knowledge & Single-item Domain Knowledge & \citet{chen2022informal} & Moderator \\
Risk Perception & Single-item Risk Perception & \citet{Holzmeister2020} & Response \\
\bottomrule
\end{tabular}
}
\end{table*}

\section{Method}
The scenario of our experiment was a financial investment scenario that was previously used in~\citet{Bao2022}, and for assessing risk perception, we borrowed the measurement protocol from~\citet{Holzmeister2020}. In the present study, users interacted with an AI agent to seek financial advice. In this scenario, the implicit consequence for the user is that if such a user invests money in the products recommended by the agent, they might gain or lose money. Thus, some level of risk is inherent in the recommendation given by the AI. From~\citet{Bao2022}, we borrowed the financial scenario, the procedure, and the design of the interface -- i.e., the three types of anthropomorphic cues which Bao et al. developed based on previous theories~\citep{seeger2021designing} -- whereas to measure risk perception we adopted a single-item test -- see~\citet{Holzmeister2020}. 

We implemented a simulated AI agent that resembled a financial advisor. The chatbot's level of anthropomorphism was manipulated by copying the design used in previous research~\citep{Seeger2021,Bao2022}. Such a design framework was borrowed from the heuristic-systematic model (HSM) found in~\citet{Bao2022}, where anthropomorphic design elements include identity cues (e.g., chatbot names signaling gender), visual cues (e.g., avatars showing the visual appearance of the AI agent), and communicative cues (speech-related elements, both verbal and non-verbal) that make the chatbot appear friendly, warm, and empathetic.

The chatbot interface and the eight manipulations proposed in~\citet{Bao2022} were also used in subsequent research -- e.g.,~\citet{sun2024anthropomorphism}, and are fully reported in Section~\ref{app:designs}. This design was chosen as multiple studies~\citep{Bao2022,sun2024anthropomorphism} confirmed its direct effect on Perceived Anthropomorphism (PA), the protagonist of this study (i.e., the predictor of trust and risk perception).      

\subsection{Chatbot Design and Experimental Conditions} \label{app:designs}

The anthropomorphic design in our study varies three cues -- visual, identity, and communicative -- within a unified framework~\citep{Bao2022} which is, in turn, an extension of~\citet{seeger2021designing}'s work. Heuristic processing relies on simple, experience-based signals; here, those signals are identity cues (e.g., names that imply gender vs. names that imply a technological artifact) and visual cues (e.g., avatars that make the agent look more human vs. more machine-like). These two categories of cues work on low-level perception and shape how users ``feel'' about the chatbot (whether it feels human-like or machine-like). Systematic processing, by contrast, engages with communicative cues -- the chatbot’s verbal style (i.e., the use of a friendly tone and small talk typical of human-to-human communication vs. the use of an analytical and distant language style communicating instructions and procedures typical of machine talk). Non-verbal cues also belong to this processing route. These consist of the manipulation of more subtle elements -- e.g., waiting time when the machine is thinking, signaled by thinking dots on the screen.    

\begin{figure*}[t]
\centering
\includegraphics[width=0.9\linewidth]{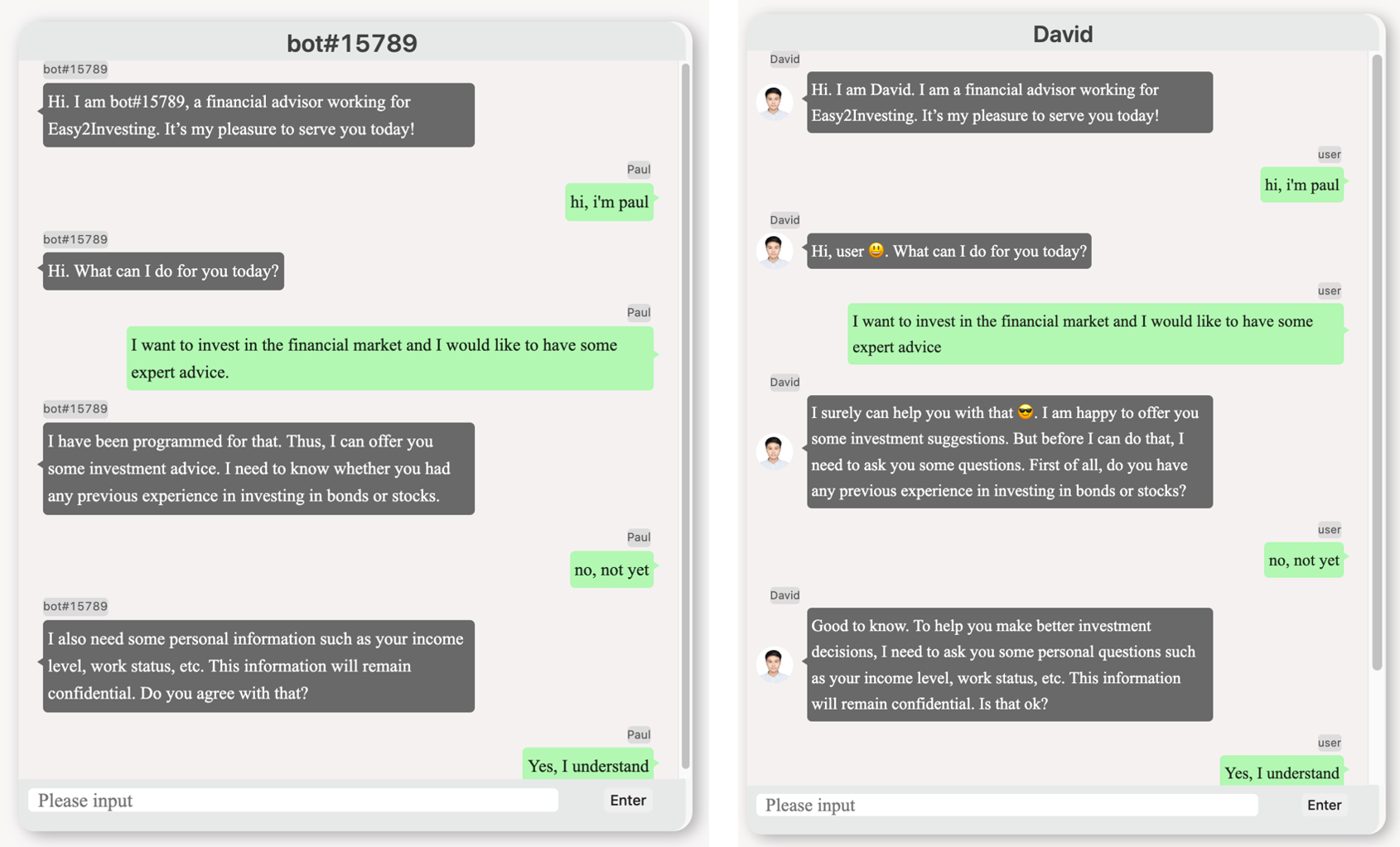}
\caption{Two side-by-side chatbot interface screenshots showing contrasting designs. The left side shows a machine-like design with bot\#15789 identifier in place of the agent name, the image of a PC in place of a human avatar, and the use of a distant and procedural-style language to communicate with the user. The right side shows the human-like design with the name of the agent (David), a human avatar photo, and anthropomorphic communication. Both show similar
conversation flow about financial investment advice with user responses in green bubbles.}
\label{fig:designs}
\end{figure*}

Operationally, we implemented a 2$\times$2$\times$2 design: Visual cue (avatar): computer image vs. human photo; Identity cue (name): ``bot\#156'' vs. ``David''; Communicative cue (tone and non-verbal elements): machine- vs. human-style communication as per~\citet{Bao2022}. This yields eight anthropomorphism conditions. The two extremes -- fully machine-like and fully human-like, are illustrated in Figure~\ref{fig:designs}. Such a manipulation replicates previous research~\citep{seeger2021designing,Bao2022,sun2024anthropomorphism} and its effect on the actual perception of anthropomorphism is rigorously assessed in Section~\ref{manipulation}, which reports the manipulation check.

\subsection{Experimental Design}
We conducted an independent between-subject design online experiment. The independent variable (IV) was Anthropomorphic Design with eight levels. The dependent variable (DV) was Perceived Anthropomorphism (PA), measured using the Godspeed questionnaire~\citep{Bartneck2009}. Then, PA was used as the main predictor of Risk Perception~\citep{Holzmeister2020} via Trust, measured using the CATQ~\citep{Johnson2005}, moderated by Domain Knowledge~\citep{chen2022informal} in both the regression and structural equation modeling -- see the theoretical framework in Section~\ref{theory}.   

\subsection{Participants}
Participants (Ps) were recruited via Prolific in exchange for a monetary reward. In the platform, we specified two eligibility criteria, such as English as the first language and a minimum education level of a high school diploma. To mitigate gender bias, the sampling was evenly split, with 50\% male and 50\% female. Ps with incomplete submissions were deleted. 

A total of 1430 people completed the study. After performing attention checks, 1256 Ps were used for further analysis (50.4\% female; mean age=38.1 years, SD = 12.9). On average, participants took 43 minutes (SD = 10.57) to complete the experiment. More than 90\% of the participants were from the US. Regarding education level, 33\% of Ps did not complete any university degree, 41\% held a bachelor’s degree, and the remaining completed a postgraduate degree. 

\subsection{Procedure}
The study was conducted on a custom web platform hosting an AI financial-advisor chatbot. Participants accessed the study via a link, reviewed the purpose of the research, and provided informed consent. They then completed a brief questionnaire collecting demographics (age, gender, education, annual income) and background information on financial knowledge and investment experience.

Participants were randomly assigned to one of eight anthropomorphic-design conditions~\citep{Bao2022}. The interface varied along three cues -- avatar (computer image vs. human photo), identity (bot-like code vs. human name), and communicative cues (machine-like vs. human-like verbal and non-verbal communication). The investment scenario and interaction flow followed the setup described in~\citet{Bao2022}.

To standardize the human-AI exchange, the interaction was guided. An instruction pane introduced the task (for example, ``Ask the advisor how to invest your savings''), and a suggestion window beside the chat offered step-by-step prompts to help participants formulate their request. Within their assigned condition, participants interacted with the chatbot, which then produced a concrete investment recommendation.

Immediately after the interaction, participants completed the study measures using 7-point Likert scales. These included Perceived Anthropomorphism of the chatbot (Godspeed items), Cognitive Trust and Affective Trust toward the advisor~\citep{Johnson2005}, self-assessed domain knowledge about financial products and investing~\citep{chen2022informal}, and perceived risk of following the AI’s recommendation~\citep{Holzmeister2020} -- see Section~\ref{variables}. All responses were recorded electronically on the platform, and participants were debriefed at the end of the session.

\section{Results}
Descriptive statistics and correlations are reported in Table~\ref{tab:correlation}. In the table, N is the variable code number. The table reports the means and standard deviations for each variable, along with the Pearson correlation coefficients between variables.    

\begin{table*}[ht]
  \centering
  \caption{Summary statistics and correlations.}
  \label{tab:correlation}
  \resizebox{\textwidth}{!}{%
  \begin{tabular}{l c c c c c c c c}
    \toprule
    \textbf{N} & \textbf{Variable} & \textbf{mean} & \textbf{std} & \textbf{1} & \textbf{2} & \textbf{3} & \textbf{4} & \textbf{5} \\
    \midrule
    1 & Cognitive Trust & 4.16 & 0.88 & 1 &  &  &  &  \\
    2 & Affective Trust & 4.09 & 1.15 & 0.45 & 1 &  &  &  \\
    3 & Perceived Anthropomorphism & 3.85 & 1.64 & 0.39 & 0.59 & 1 &  &  \\
    4 & Risk Perception & 3.57 & 1.59 & -0.12 & 0.08 & 0.10 & 1 & \\
    5 & Domain Knowledge & 3.96 & 1.67 & 0.11 & 0.33 & 0.26 & 0.17 & 1 \\
    \bottomrule
  \end{tabular}
  }
\end{table*}

\subsection{Testing the Design Manipulation}\label{manipulation} 
To test whether the designs adopted affected the participants' Perceived Anthropomorphism, the actual predictor in our model, measured using the Godspeed questionnaire developed by Bartneck et al.~\citep{Bartneck2009} -- see Figure~\ref{fig:moderated}, we conducted ANOVA and Post Hoc testing. 

A one-way ANOVA shows that the Anthropomorphic Design adopted had a significant effect on Perceived Anthropomorphism -- F(7, 1248) = 2.8912, p = 0.005, $\eta^2$ = 0.016. This confirms that the 8-level design functions as an antecedent of Perceived Anthropomorphism, in line with previous results~\citep{Bao2022} -- i.e., anthropomorphic design manipulation directly affects Perceived Anthropomorphism. 

Tukey HSD (alpha = .05) revealed that the most human-like bundle (human photo + human name + human-like communication) exceeded a mix design (human photo + human name + machine-like communication) -- $\Delta$ = 0.650, p-adj = .010, Cohen’s d = 0.405 (moderate), and the fully machine-like design (machine photo + machine name + machine-like communication -- $\Delta$ = 0.572, p-adj = .026, Cohen’s d = 0.350 (moderate). Other pairwise differences were non-significant.

After testing marginal contrasts of the design (flipping one cue while averaging over the other two) with Welch t-tests and Holm correction, we found the results displayed in Table~\ref{tab:cues}, confirming the stability of our manipulation.

\begin{table*}[ht]
  \centering
  \caption{Marginal contrasts for anthropomorphic design cues on Perceived Anthropomorphism.}
  \label{tab:cues}
  \resizebox{\textwidth}{!}{%
  \begin{tabular}{@{}l p{0.32\linewidth} c c c c c c@{}}
    \toprule
    \textbf{Cue} &
    \textbf{Levels (A - B)} &
    \textbf{Mean$_A$} &
    \textbf{Mean$_B$} &
    \boldmath$\Delta$ &
    \textbf{t} &
    \textbf{p} &
    \textbf{Effect size} \\
    \midrule
    Communication & human-like vs machine-like        & 3.996 & 3.706 & 0.290 & 3.141 & .005 & 0.177 / 0.177 \\
    Visual        & human photo vs computer image     & 3.953 & 3.749 & 0.204 & 2.296 & .050 & 0.127 / 0.126 \\
    Identity      & human name vs bot code            & 3.859 & 3.852 & 0.007 & 0.083 & .934 & 0.005 / 0.005 \\
    \bottomrule
  \end{tabular}
  }
\end{table*}

In summary, the 8-level design does alter Perceived Anthropomorphism overall, and the post-hoc pattern pinpoints communicative cues as the key lever: making the AI agent write in a human-like manner boosts Perceived Anthropomorphism. Using an avatar (a human photo) contributes a smaller, borderline increase, whereas using a human name (vs a bot code) is inert on its own. The highest Perceived Anthropomorphism is achieved by the fully human-like condition (human photo + human name + human-like communication), which significantly outperforms comparable variants with machine-like communication style, underscoring that tone carries the heaviest weight among the three cues. Once again, these results are in line with previous research showing a similar effect of AI design on Perceived Anthropomorphism~\citep{Bao2022}. 

\subsection{Anthropomorphic Design on Trust}\label{manipulatrust} 

We also performed a supplementary cue-combination analysis on the mediator Trust. Although the 2$\times$2$\times$2 anthropomorphic design reliably shifted Perceived Anthropomorphism (PA), the same cue combinations did not produce direct differences in trust. A factorial cue model (Visual x Identity x Communication) predicting Cognitive Trust (CT) and Affective Trust (AT) showed no significant main effects or interactions on either trust dimension (CT: all $ps \ge $.071; AT: all $ps \ge $.082), and treating the manipulation as an 8-level condition factor likewise yielded non-significant omnibus effects (CT: F(7,1248)=0.78, p = .604; AT: F(7,1248)=1.25, p = .275). Conversely, Section~\ref{regress} below shows that Perceived Anthropomorphism strongly predicts trust in the main model. This pattern suggests that the experimental cues primarily operate by shaping participants' psychological interpretation of the agent's human-likeness (PA), rather than directly shifting trust as an outcome on their own. Previous research has shown that people vary significantly in their baseline tendency to anthropomorphize non-human entities, with these individual differences linked to personality, cognitive styles, demographic factors, and even neural structure, implying that different users respond very differently to the same anthropomorphic cues~\citep{waytz_cacioppo_epley_2010_who_sees_human,letheren_kuhn_lings_pope_2016_anthropomorphic_tendency}. In our analysis, we treat Anthropomorphic Design as the cause of Perceived Anthropomorphism after verifying its manipulation (i.e., the results in Section~\ref{manipulation} serve as manipulation checks). Then, we use Perceived Anthropomorphism as the main predictor in the conditional process model because it captures the effective dose of anthropomorphism experienced by participants—integrating the manipulated cues with individual differences in how those cues are perceived; whereas the design conditions are a more distal manipulation whose downstream effects on trust are largely indirect via Perceived Anthropomorphism.

\subsection{Perceived Anthropomorphism on Risk Perception Via Trust}\label{regress} 

In the following analyses, we model Perceived Anthropomorphism as the proximal psychological driver of Risk Perception via Trust.  

The regression analysis is shown in Table~\ref{tab:ols_models} (Hierarchical OLS for RP) and Table~\ref{tab:mediators} (Mediator Models). See also Equation \ref{eq:regression}, which describes the regression model.

\begin{equation} 
\label{eq:regression}
\begin{aligned} 
RP_i =\;& \beta_0 + \beta_1 \text{Age}_i + \beta_2 \text{Gender}_i + \beta_3 \text{Education}_i \\ 
&+ \beta_4 \text{Income}_i + \beta_5 \text{PA}_i + \beta_6 \text{CT}_i + \beta_7 \text{AT}_i \\
&+ \beta_8 \text{DK}_i + \beta_9 \text{PA}_i \times \text{DK}_i + \epsilon_i
\end{aligned}
\end{equation}

\begin{table*}[ht]
  \centering
  \caption{Dependent variable = Risk Perception (Hierarchical OLS Models).}
  \label{tab:ols_models}
  \resizebox{\textwidth}{!}{%
  \begin{tabular}{l*{6}{c}}
    \toprule
    \textbf{Variable} & \textbf{Model 0} & \textbf{Model 1} & \textbf{Model 2} & \textbf{Model 3} & \textbf{Model 4} & \textbf{Model 5} \\
    \midrule
    Intercept & 3.307$^{***}$ & 3.087$^{***}$ & 4.204$^{***}$ & 4.085$^{***}$ & 3.909$^{***}$ & 4.348$^{***}$ \\
    Age & $-0.003$ & $-0.004$ & $-0.004$ & $-0.003$ & $-0.003$ & $-0.002$ \\
    Gender & $-0.218^{*}$ & $-0.213^{*}$ & $-0.198^{*}$ & $-0.184^{*}$ & $-0.111$ & $-0.107$ \\
    Education & 0.204$^{***}$ & 0.177$^{**}$ & 0.147$^{**}$ & 0.129$^{*}$ & 0.064 & 0.047 \\
    Income & 0 & 0 & 0 & 0 & 0 & 0 \\
    Perceived Anthropomorphism & & 0.084$^{**}$ & 0.154$^{***}$ & 0.120$^{**}$ & 0.112$^{**}$ & $-0.021$ \\
    Cognitive Trust & & & $-0.320^{***}$ & $-0.359^{***}$ & $-0.360^{***}$ & $-0.354^{***}$ \\
    Affective Trust & & & & 0.105$^{*}$ & 0.075 & 0.080 \\
    Domain Knowledge & & & & & 0.114$^{***}$ & $-0.010$ \\
    PA$\times$DK & & & & & & 0.032 \\
    \midrule
    $R^2$ & 0.023 & 0.030 & 0.057 & 0.060 & 0.070 & 0.073 \\
    \bottomrule
  \end{tabular}
  }
  \vspace{0.5em}
  \makebox[\linewidth][l]{\parbox{\linewidth}{\textbf{Note.} PA $\times$ DK = Interaction term between Perceived Anthropomorphism and Domain Knowledge; No star: Not significant ($p \ge 0.05$), * : $p < 0.05$, ** : $p < 0.01$, and *** : $p < 0.001$.}}
\end{table*}

The base model (Model 0) shows that Age, gender, education, and income (included as controls) have a modest effect ($R^2= 0.023$) and explain only a small portion of variance in RP. Adding Perceived Anthropomorphism (PA) in Model 1 improves model fit ($R^2 = .030$) and supports H1: higher Perceived Anthropomorphism is associated with higher Risk Perception. The mediation models then examine how Cognitive Trust (CT) and Affective Trust (AT) transmit the effect of PA on RP, and how Domain Knowledge (DK) moderates these links.

For completeness, we also report the distribution of affective trust (AT) across different levels of Perceived Anthropomorphism (PA) in~\ref{app1}. 

Table~\ref{tab:hypothesis_tests} summarizes the core hypothesis tests across models.

\begin{table*}[h]
  \centering
  \caption{Summary of hypothesis tests from regression models.}
  \label{tab:hypothesis_tests}
  \resizebox{\textwidth}{!}{%
  \begin{tabular}{lllcccc}
    \toprule
    \textbf{Hypothesis} & \textbf{Path / Effect} & \textbf{Model} & $R^2$ & $b$ & $p$ & \textbf{Supported} \\
    \midrule
    H1   & PA $\rightarrow$ RP (direct) & Model 1   & .030 & 0.084    & .0068   & Yes \\
    H2A  & PA $\rightarrow$ CT          & CT M1     & .160 & 0.218    & $<$.001 & Yes \\
    H2B  & CT $\rightarrow$ RP (controlling PA) & Model 2 & .057 & $-0.320$ & $<$.001 & Yes \\
    H3A  & PA $\rightarrow$ AT          & AT M1     & .367 & 0.401    & $<$.001 & Yes \\
    H3B  & AT $\rightarrow$ RP (controlling PA \& CT) & Model 3 & .060 & 0.105 & .048   & Yes \\
    H4   & DK $\rightarrow$ RP          & Model 4   & .070 & 0.114    & $<$.001 & Yes \\
    H4a  & PA$\times$DK $\rightarrow$ CT & CT M3    & .168 & $-0.0228$ & .0086  & Yes \\
    H4b  & PA$\times$DK $\rightarrow$ AT & AT M3    & .390 & $-0.0253$ & .0071  & Yes \\
    \bottomrule
  \end{tabular}
  }
  \vspace{0.5em}
  \makebox[\linewidth][l]{\parbox{\linewidth}{\textbf{Note.} PA = Perceived Anthropomorphism; CT = Cognitive Trust; AT = Affective Trust; DK = Domain Knowledge; RP = Risk Perception; PA$\times$DK = interaction term between Perceived Anthropomorphism and Domain Knowledge; $b$ = unstandardized regression coefficient; $R^2$ = proportion of variance in the dependent variable explained by the model. (Base Model 0 with controls only: $R^2 = .023$, not shown in table.)}}
\end{table*}

Taken together, these results indicate that Perceived Anthropomorphism has a small positive direct association with Risk Perception (H1), but also exerts an indirect, risk-reducing effect via Cognitive Trust: higher Perceived Anthropomorphism increases Cognitive Trust (H2A), and in turn, higher Cognitive Trust predicts lower Risk Perception (H2B). Perceived Anthropomorphism likewise increases Affective Trust (H3A), yet Affective Trust shows a small positive association with Risk Perception once Perceived Anthropomorphism and Cognitive Trust are controlled (H3B), suggesting that Affective Trust follows a different route from Cognitive Trust. Finally, Domain Knowledge both raises Risk Perception directly (H4) and weakens the impact of Perceived Anthropomorphism on Cognitive Trust and Affective Trust (H4a, H4b): knowledgeable participants are less swayed by anthropomorphic cues in terms of both competence-related and Affective Trust.

\begin{table*}[ht]
  \centering
  \caption{Mediator models for Cognitive and Affective Trust.}
  \label{tab:mediators}
  \resizebox{\textwidth}{!}{%
  \begin{tabular}{l cccc cccc}
    \toprule
    & \multicolumn{4}{c}{DV: Cognitive Trust} & \multicolumn{4}{c}{DV: Affective Trust} \\
    \cmidrule(lr){2-5} \cmidrule(lr){6-9}
    \textbf{Variable} & \textbf{CT M0} & \textbf{CT M1} & \textbf{CT M2} & \textbf{CT M3} & \textbf{AT M0} & \textbf{AT M1} & \textbf{AT M2} & \textbf{AT M3} \\
    \midrule
    Intercept & 4.055*** & 3.486*** & 3.407*** & 3.068*** & 3.457*** & 2.412*** & 2.161*** & 1.784*** \\
    Age & 0.003 & 0 & 0 & 0 & 0.002 & -0.004 & -0.003 & -0.004* \\
    Gender & 0.036 & 0.048 & 0.072 & 0.069 & -0.129* & -0.108* & -0.032 & -0.035 \\
    Edu & -0.024 & -0.095** & -0.116*** & -0.105*** & 0.264*** & 0.134*** & 0.065 & 0.078* \\
    Income & 0 & 0 & 0 & 0 & 0.0** & 0.0** & 0 & 0 \\
    Perceived Anthro. & & 0.218*** & 0.211*** & 0.303*** & & 0.401*** & 0.380*** & 0.482*** \\
    Domain Knowledge & & & 0.035* & 0.123** & & & 0.112*** & 0.210*** \\
    PA$\times$DK & & & & -0.023** & & & & -0.025** \\
    \midrule
    $R^2$ & 0.004 & 0.160 & 0.163 & 0.168 & 0.058 & 0.367 & 0.386 & 0.390 \\
    \bottomrule
  \end{tabular}
  }
  \vspace{0.5em}
\makebox[\linewidth][l]{\parbox{\linewidth}{\textbf{Note.} PA$\times$DK = Interaction term between Perceived Anthropomorphism and Domain Knowledge; No star: Not significant ($p \ge 0.05$), * : $p < 0.05$, ** : $p < 0.01$, and *** : $p < 0.001$.}}
\end{table*}

We further examined Domain Knowledge as a moderator using conditional analyses at three levels (low = mean $- 1$ SD = 2.29; medium = 3.96; high = mean $+ 1$ SD = 5.63). 

\begin{figure*}[t]
\centering
\includegraphics[width=0.7\linewidth]{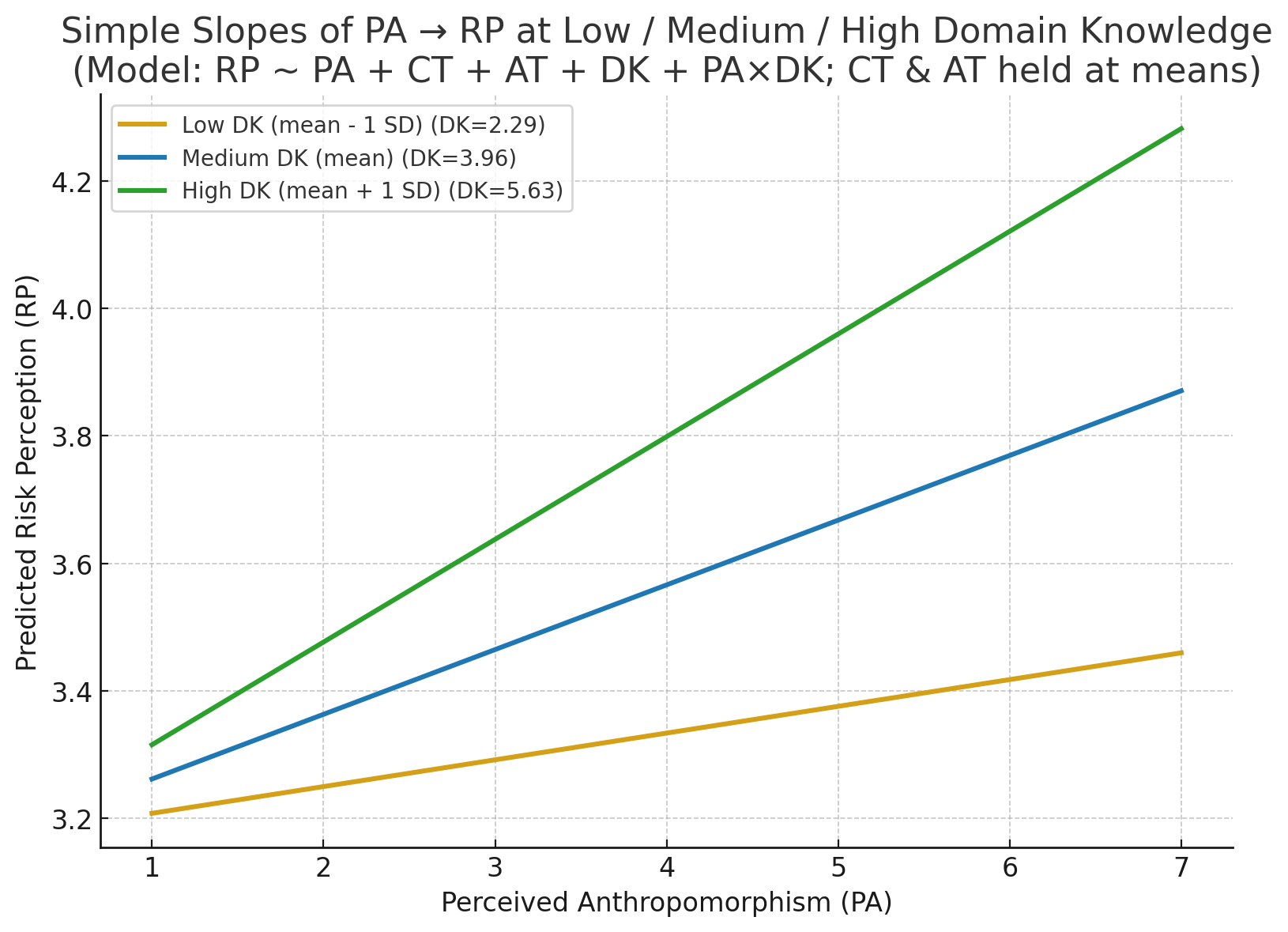}
\caption{A line graph showing three upward-sloping lines representing the relationship between Perceived Anthropomorphism on the x-axis and
Predicted Risk Perception on the y-axis. The yellow line shows low Domain Knowledge (2.29), the blue line shows medium Domain Knowledge (3.96), and
the green line shows high Domain Knowledge (5.63). The green line has the steepest slope, indicating a stronger relationship at high Domain Knowledge.}
\label{fig:slope}
\end{figure*}

The indirect effect via Cognitive Trust was negative and significant at low Domain Knowledge ($b = -0.093$, 95\% CI [$-0.128, -0.059$]), and it weakened (i.e., became less negative) but remained significant at high Domain Knowledge ($b = -0.060$, 95\% CI [$-0.087, -0.037$]). The direct effect of Perceived Anthropomorphism on Risk Perception increased with Domain Knowledge: at low DK the simple slope was 0.042 (95\% CI [$-0.057, 0.141$], n.s.); at medium DK it was 0.101 (95\% CI [$0.029, 0.174$], $p = .0065$); and at high DK it was 0.161 (95\% CI [$0.070, 0.248$], $p < .001$). A simple-slopes plot (Figure~\ref{fig:slope}) illustrates the progressively steeper Perceived Anthropomorphism$\rightarrow$ Risk Perception slope as Domain Knowledge rises. Steeper lines indicate that perceived anthropomorphism has a stronger positive association with risk perception at that domain-knowledge level; the increasing steepness from Low to High DK visualizes the positive  PA $\times$ DK interaction.

Moderated-mediation analyses confirmed that the CT-mediated pathway depends on DK: the index of moderated mediation (CT path) was positive and significant (index = 0.0096, 95\% CI [$0.0034, 0.0167$]), indicating that the (negative) indirect effect via CT attenuates as DK increases.

\begin{figure*}[t]
\centering
\includegraphics[width=0.7\linewidth]{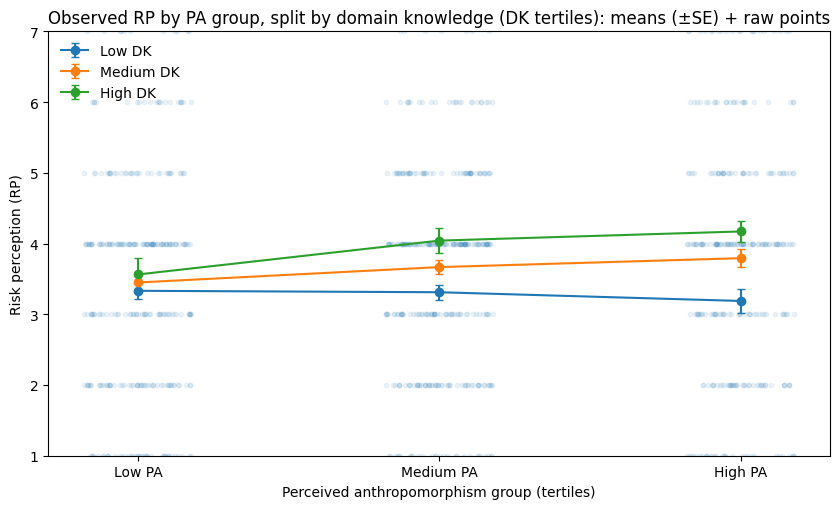}
\caption{PA is discretised into tertiles (Low, Medium, and High PA) and plotted on the x-axis. Lines show observed mean RP within each PA group for low, medium, and high DK (tertiles), with standard error bars. Light points represent raw participant responses (jittered for visibility). This figure visualises the underlying observed data pattern used to interpret the  PA $\times$ DK moderation, rather than model-predicted slopes.}
\label{fig:interaction}
\end{figure*}

For completeness, Figure~\ref{fig:interaction} visualises raw Risk Perception responses (RP) as a function of Perceived Anthropomorphism (PA), stratified by Domain Knowledge (DK). PA is grouped into tertiles (low, medium, and high PA) to provide an interpretable, distribution-level view of the data; within each PA group, the lines represent observed mean RP for participants with low, medium, and high DK (tertiles), with error bars indicating the standard errors, and the jittered points showing individual observations. The plot indicates that Risk Perception varies systematically across Perceived Anthropomorphism levels and that this pattern differs by Domain Knowledge: the separation between DK groups becomes more pronounced as PA increases, consistent with a  PA $\times$ DK interaction in which higher-knowledge participants appear more sensitive to variations in perceived anthropomorphism when judging investment risk. Importantly, this figure complements the inferential models by showing the empirical variability and dispersion underlying the moderation results, rather than relying solely on model-predicted simple slopes. 

To further check the robustness of our model and explore other potential interactions between variables, we additionally tested PA$\times$CT and PA$\times$AT interactions predicting RP. Both were marginal and did not alter the main mediation conclusions. This analysis can be found in~\ref{app2}. 

In summary, the regression analysis shows that Perceived Anthropomorphism (PA) directly raises Risk Perception (RP) (H1 supported) and indirectly lowers RP via Cognitive Trust (CT) (because PA increases CT and, in turn, CT reduces RP), creating opposing pathways (H2). PA also raises Affective Trust (AT), and AT (controlling PA \& CT) slightly raises RP (H3b), suggesting that Affective Trust may track warmth/liking without conferring a sense of safety. Domain Knowledge heightens Risk Perception overall (H4) and attenuates the impact of anthropomorphic cues on both Cognitive Trust and Affective Trust (H4a/H4b).

\subsection{SEM Analysis}
\begin{figure*}[t]
\centering
\includegraphics[width=0.7\linewidth]{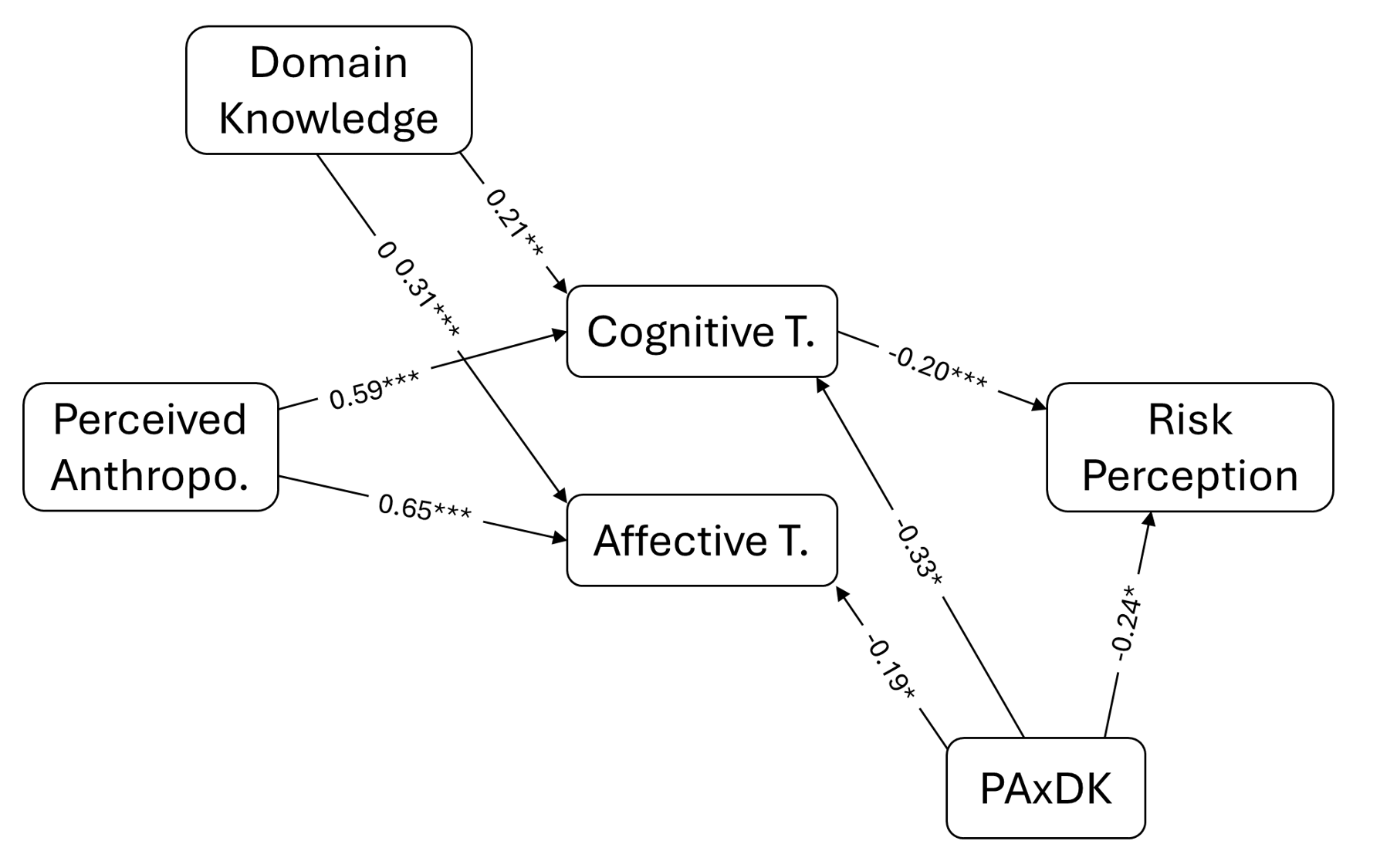}
\caption{Path Diagram for Structural Equation Model. Only significant coefficients (and paths) are represented. The variables are Perceived Anthropomorphism, Cognitive Trust, Affective Trust, Domain Knowledge, Risk Perception; and PA$\times$DK = interaction term between Perceived Anthropomorphism and Domain Knowledge.}
\label{fig:sem}
\end{figure*}

We also performed structural equation modelling (SEM). Figure~\ref{fig:sem} shows these results with the coefficients for the significant paths. In the SEM, Perceived Anthropomorphism (PA) strongly predicted both forms of trust: AT ($\beta = 0.653, p < .001$) and CT ($\beta = 0.586, p < .001$). Domain Knowledge (DK) also positively predicted trust---AT ($\beta = 0.308, p < .001$) and CT ($\beta = 0.214, p = .002$)---but attenuated the impact of PA on both trusts, as indicated by significant PA$\times$DK interactions for AT ($\beta = -0.187, p = .034$) and CT ($\beta = -0.330, p = .002$). Downstream, CT was negatively associated with Risk Perception (RP) ($\beta = -0.201, p < .001$), whereas the direct PA$\rightarrow$RP and AT$\rightarrow$RP paths were not significant in this SEM. We did observe a small positive PA$\times$DK$\rightarrow$RP path ($\beta = 0.243, p = .049$), suggesting that higher DK can amplify the (direct) association between PA and Risk Perception. Model fit (variance explained) was modest for RP ($R^2 = .071$) and larger for the trust mediators ($R^2_{AT} = .384$; $R^2_{CT} = .159$), consistent with a mechanism in which anthropomorphic cues primarily shape trust (especially Affective Trust), while Cognitive Trust in turn reduces Risk Perception. For the full SEM analysis with all the paths and their coefficients, see Section~\ref{app:sem_results}.

\subsubsection{Structural Equation Model Path Analysis Results} \label{app:sem_results}

Here, we report the SEM analysis with all the paths, the coefficients, p-values, and significance levels. Table~\ref{tab:sem_paths} presents the complete results of the structural equation model path analysis.

\begin{table*}[htbp]
\centering
\caption{SEM Path Analysis Results: Standardized Coefficients and Significance Tests.}
\label{tab:sem_paths}
\resizebox{\textwidth}{!}{%
\begin{tabular}{@{}llrrrl@{}}
\toprule
\textbf{Dependent Variable} & \textbf{Independent Variable} & \textbf{Std. Coeff.} & \textbf{p-value} & \textbf{Sig.} \\
\midrule
\multirow{7}{*}{Affective Trust} & Domain Knowledge & 0.304 & 0.000 & *** \\
                    & Education & 0.057 & 0.025 & * \\
                    & Gender & -0.015 & 0.500 & \\
                    & Perceived Anthropomorphism & 0.686 & 0.000 & *** \\
                    & PA$\times$DK & -0.240 & 0.007 & ** \\
                    & Age & -0.045 & 0.037 & * \\
                    & Income & 0.035 & 0.151 & \\
\midrule
\multirow{7}{*}{Cognitive Trust} & Domain Knowledge & 0.234 & 0.001 & ** \\
                    & Education & -0.099 & 0.001 & *** \\
                    & Gender & 0.039 & 0.136 & \\
                    & Perceived Anthropomorphism & 0.565 & 0.000 & *** \\
                    & PA$\times$DK & -0.282 & 0.009 & ** \\
                    & Age & -0.006 & 0.828 & \\
                    & Income & -0.012 & 0.484 & \\
\midrule
\multirow{9}{*}{Risk Perception} & Affective Trust & 0.058 & 0.139 & \\
                    & Cognitive Trust & -0.195 & 0.000 & *** \\
                    & Domain Knowledge & -0.011 & 0.894 & \\
                    & Education & 0.025 & 0.423 & \\
                    & Gender & -0.034 & 0.235 & \\
                    & Perceived Anthropomorphism & -0.021 & 0.807 & \\
                    & PA$\times$DK & 0.218 & 0.084 & \\
                    & Age & -0.019 & 0.491 & \\
                    & Income & 0.023 & 0.500 & \\
\bottomrule
  \end{tabular}
  }
  \vspace{0.5em}
\makebox[\linewidth][l]{\parbox{\linewidth}{\textbf{Note.} PA$\times$DK = Interaction term between Perceived Anthropomorphism and Domain Knowledge; No star: Not significant ($p \ge 0.05$), * : $p < 0.05$, ** : $p < 0.01$, and *** : $p < 0.001$.}}
\end{table*}

In summary, from the analysis we observed that:
\begin{itemize}
    \item Perceived Anthropomorphism positively and significantly affects Cognitive Trust (supports H2A).
    \item Cognitive Trust negatively and significantly affects Risk Perception (this supports H2B; mediation via Cognitive Trust).
    \item Perceived Anthropomorphism positively and significantly affects Affective Trust (supports H3A).
    \item Affective Trust positively and significantly affects RP, although the effect is small (supports H3B, direction positive).
    \item Perceived Anthropomorphism positively and significantly affects Risk Perception, and this effect is also small (supports H1 direct).
    \item Domain Knowledge positively and significantly affects Risk Perception (supports H4).
    \item The interaction Perceived Anthropomorphism $\times$ Domain Knowledge affects both Cognitive Trust and Affective Trust negatively and significantly (supports H4a, H4b: DK attenuates PA's effect on both trusts).
\end{itemize}

\subsection{Robustness Checks}
Multicollinearity is not a substantive concern in our models. The only structurally collinear term is the interaction Perceived Anthropomorphism $\times$ Domain Knowledge, which (by construction) correlates with its components. Nevertheless, to further test the robustness of our results, we performed a multicollinearity test. As standard practice, we also mean-centered Perceived Anthropomorphism and Domain Knowledge before forming the interaction.

\subsubsection{Multicollinearity test}
We assessed multicollinearity using variance inflation factors (VIFs) and condition-index diagnostics across all models. In the risk-perception equation, uncentered predictors produced the following VIFs: Perceived Anthropomorphism = 7.29, Cognitive Trust = 1.31, Affective Trust = 1.79, Domain Knowledge = 6.66, Interaction = 14.68. After mean-centering, Perceived Anthropomorphism and Domain Knowledge and re-forming the interaction, VIFs dropped to: centered Perceived Anthropomorphism = 1.60, Cognitive Trust = 1.31, Affective Trust = 1.79, centered Domain Knowledge = 1.14, centered Interaction = 1.02. The corresponding condition-index spectrum for the centered model was 1.00, 1.41, 1.45, 1.58, 1.90, 2.32.

For the mediator specifications (Cognitive Trust and Affective Trust each regressed on Perceived Anthropomorphism, Domain Knowledge, and their interaction), uncentered VIFs were: Perceived Anthropomorphism = 6.43, Domain Knowledge = 6.49, Interaction = 14.53; after centering: centered Perceived Anthropomorphism = 1.07, centered Domain Knowledge = 1.08, centered Interaction = 1.01. The condition-index spectrum for the centered mediator models was 1.00, 1.14, 1.16, 1.32. These diagnostics show that the apparent inflation is the expected artifact of product terms before centering and that centering removes nonessential collinearity to benign levels -- see~\citet{aiken1991multiple}.

\subsubsection{Refitting with centered predictors}
Using the centered specification from the preceding subsection, we refit the models and report only substantive effects. In the ordinary least squares model for Risk Perception, adding Perceived Anthropomorphism to Domain Knowledge yielded a positive Perceived Anthropomorphism $\rightarrow$ Risk Perception effect (b = 0.063, SE = 0.031, p = .042). With mediators included, Cognitive Trust was negative and reliable (b = -0.230, SE = 0.030, $p<.001$), Affective Trust was small positive (b = 0.092, SE = 0.046, p = .045), and Domain Knowledge remained positive (b = 0.134, SE = 0.028, $p<.001$); the centered interaction between Perceived Anthropomorphism and Domain Knowledge was modest (b = 0.073, SE = 0.039, p = .056; model R-sq = .071). Standardized path models reproduced these patterns: Perceived Anthropomorphism $\rightarrow$ Affective Trust (b = 0.596, $p<.001$) and Perceived Anthropomorphism $\rightarrow$ Cognitive Trust (b = 0.364, $p<.001$) were positive; Domain Knowledge $\rightarrow$ Affective Trust (b = 0.325, $p<.001$) and Domain Knowledge $\rightarrow$ Cognitive Trust (b = 0.210, $p<.001$) were positive; the interaction between Perceived Anthropomorphism and Domain Knowledge attenuated the effects on Affective Trust (b = -0.180, $p<.001$) and Cognitive Trust (b = -0.133, $p<.001$). For Risk Perception, Cognitive Trust was negative (b = -0.239, $p<.001$), Affective Trust was small positive (b = 0.091, p = .004), the interaction effect on Risk Perception was modest (b = 0.064, p = .049), and the direct effect of Perceived Anthropomorphism on Risk Perception was non-significant once mediators were included (b = 0.025, p = .346) -- collinearity diagnostics for the centered specification are provided in the previous subsection.

After diagnosing and addressing multicollinearity (mean-centering Perceived Anthropomorphism and Domain Knowledge before forming their interaction), all coefficients and inference were stable. The ordinary least squares models with robust errors and the standardized, observed-variable SEM converged on the same pattern: Perceived Anthropomorphism increases both Affective Trust (strongly) and Cognitive Trust (moderately); Domain Knowledge also increases both trusts, while the Perceived Anthropomorphism × Domain Knowledge interaction attenuates these trust gains. Downstream, Cognitive Trust reliably reduces Risk Perception, Affective Trust exerts a small positive effect on Risk Perception, and the direct path from Perceived Anthropomorphism to Risk Perception becomes non-significant once mediators are included~\citep{belsley2005regression}. In short, after centering and formal collinearity checks, the regression modeling and SEM results remain valid, and our conditional process (moderated-mediation) interpretation holds.

\section{Discussion}
Our study set out to explain how Perceived Anthropomorphism predicts Risk Perception through two forms of trust—Cognitive Trust and Affective Trust—and how Domain Knowledge conditions these links.
Most of the prior research has studied the effect of anthropomorphism on trust or on risk perception. These two mental processes are seldom discussed together in the area of AI.  Moreover, moderation by expertise is under-examined, and evidence from high-stakes decision support (e.g., finance) remains limited. We address these gaps by (i) treating anthropomorphic design as the antecedent of anthropomorphism -- e.g.,~\citet{Bartneck2009,Bao2022,sun2024anthropomorphism}. Thus, our goal is not to attribute effects to specific cues, but to understand how users' experienced human-likeness relates to trust and risk judgments. This strategy limits the drawbacks and interferences produced by certain design styles~\citep{Aubel2022,cohn2024believing} and individual differences in cue responses~\citep{waytz_cacioppo_epley_2010_who_sees_human,letheren_kuhn_lings_pope_2016_anthropomorphic_tendency}. In this work, we modelled two trust routes~\citep{Johnson2005} between anthropomorphism and Risk Perception, testing Domain Knowledge as a moderator in a large sample to account for the discrepancies found in the literature (see section~\ref{relwork}).

We posited a direct effect of anthropomorphism on Risk Perception without specifying direction, given mixed prior findings~\citep{Kim2011,Aubel2022}. The result of the regression modelling suggests that anthropomorphism positively predicted Risk Perception, but in the observed-variable SEM (excluding controls), this direct path was not retained once mediators entered—indicating that anthropomorphism influence is largely indirect, via trust, when the full network is modeled. Crucially, Domain Knowledge conditions this relation: higher expertise steepens the effect that anthropomorphism has on Risk Perception.
The mediated indirect effect of anthropomorphism on Risk Perception via Cognitive Trust was more negative at low Domain Knowledge compared to high Domain Knowledge. Moreover, the risk-reducing pathway via Cognitive Trust attenuates as Domain Knowledge increases. In short, anthropomorphism reduces Risk Perception via CT, especially among less knowledgeable users.
Moreover, the effect of Affective Trust on Risk Perception was positive, but small. In the reduced SEM, it was not retained once all paths/interactions were estimated. Overall, after accounting for Cognitive Trust, greater Affective Trust does not reliably lower RP and can be slightly risk-raising (although this latter effect was not significant)—consistent with the idea that warmth without competence may not assure safety in high-stakes advice.
Finally, Domain Knowledge does not appear to directly predict Risk Perception once mediators and interactions were included; instead, its influence emerged indirectly and via moderation. The results show that expert users are less swayed (both cognitively and effectively) by anthropomorphic cues. Furthermore, we found a steeper effect of anthropomorphism on Risk Perception for users with higher knowledge of finance. Although the results of our analyses are robust, it is worth noting that they suggest but do not prove that trust, per se, reduces risk perception. However, at the very least, these results show that trust alters how users subjectively appraise the potential consequences of AI advice.

In summary, in our theoretical model and results, anthropomorphism shifts both competence inferences and perceived warmth. In high-risk contexts, these routes appear to push Risk Perception in opposite directions: competence dampens it, whereas warmth—controlling for competence—need not. Our results fit this dual-route account and clarify the role of expertise. 

To conclude, anthropomorphism reduces risk perception via trust, so that users may feel safer with an AI that is actually capable of producing severe factual errors. This mismatch of perceived safety and actual risk echoes the recent findings on users overtrusting AI-generated content and the propagation of epistemic error~\citep{klingbeil2024trust,shekar2025people}. Our empirical evidence suggests that anthropomorphism is not just a fun or engagement-boosting feature. It actually shapes risk perception via trust. This supports a more cautious design proposal for AI system interfaces in high-risk tasks -- such systems should use anthropomorphic AI responsibly, attenuating the likelihood of users accepting hallucinated or incorrect outputs without scrutiny.

\subsection{Theoretical Contribution} 
Our model reconciles an apparent contradiction in the literature: anthropomorphism can increase trust and yet be associated with higher Risk Perception in high-stakes contexts. We find two concurrent mechanisms that were, to our knowledge, unexplored in the research on anthropomorphic AI. Anthropomorphism increases both dimensions of trust. Then, the cognitive route (competence) yields a risk-reducing indirect effect, stronger for laypeople. This comes as no surprise. Slovic's research and the psychometric paradigm of risk perception anticipated such a relationship already four decades ago~\citep{Slovic1985,Slovic2005}, although not in response to anthropomorphism and not in the context of AI. Conversely, the affective route (warmth) seems to increase Risk Perception -- this effect was significant in the regression model but non-significant in the SEM model, and can have some explanation in the Uncanny Valley effect and some Behavioral Economics theories. Under the Uncanny Valley lens~\citep{mori2012uncanny}, human-like language without fully human coherence can feel "almost-but-not-quite" human, triggering unease. In high-stakes finance, that unease is read as higher uncertainty/risk, nudging Risk Perception upward. Under the behavioral economics lens~\citep{loewenstein2001risk}, emotional tone can signal lower rationality (more noise, bias, "selling" rather than evidence). Users infer less calculative rigor, elevating Risk Perception, especially when competence cues aren't salient. These effects are stronger in high-stakes contexts and among experts, who penalize warmth more and show a steeper PA$\rightarrow$RP slope. Furthermore, the net effect of anthropomorphism depends on the user's expertise: novices benefit more from the cognitive-mediated, risk-dampening route; experts show attenuated Cognitive Trust mediation and heightened sensitivity to risk as Perceived Anthropomorphism increases. In summary, we extend human–AI interaction models by incorporating a dual-process trust mechanism linking anthropomorphism to Risk Perception, moderated by expertise, and partially explain some mixed effects reported in the literature -- e.g., ~\citet{Kim2011,Aubel2022}.

\subsection{Design Implications and Prescriptions} \label{desing}

Our results also have crucial implications for AI design. First, the manipulation of language (tone and non-verbal cues), rather than the use of human names or avatars, seems to be the real driver of perceived anthropomorphism. Moreover, our results suggest that, in financial decision support, designers could optimize anthropomorphism for Cognitive Trust: competence-forward cues (explanations, calibration, provenance, clarity) may lower Risk Perception. Affective styling alone (friendliness, small talk) is not sufficient and may even backfire once competence is accounted for, especially with expert users who may be less influenced by anthropomorphic cues and more sensitive to risk. Personalizing the degree and type of anthropomorphism to user expertise could be an option. The design takeaway is that competence-forward cues (transparency, rationale, uncertainty disclosure) may help reduce Risk Perception, if this is a sought-after effect; for example, such cues may be used for persuading users to take (hopefully ethical) actions. The enhancement of warmth in human-AI interaction (generally associated with high anthropomorphism) may be used sparingly and contextually so it complements, rather than dilutes, signals of analytical rigor. Given the growing pervasiveness of AI, these findings should inform the responsible design of such systems. 


To make these implications actionable, we propose three concrete examples of system design that could shape perceived risk and its interplay with perceived human-likeness, trust, and user expertise: (1) A competence-first frame: designers could pair every recommendation with brief calibration and provenance (what data, how validated, with what uncertainty) and a short, plain-language rationale. This may strengthen judgments of competence and reduce perceived risk when appropriate. (2) Gate tone by evidence and sequence warmth after competence: friendly language appears only alongside calibration or rationale, so warmth complements, rather than replaces, signals of analytical rigor. (3) Adapt to user expertise: developers could collect a quick self-assessment of knowledge at onboarding, then present more formal, evidence-heavy messages to knowledgeable users and more guided, explanatory messages to novices; they could also add guardrails so that for knowledgeable users, the interface automatically down-weights warmth and surfaces uncertainty. When needed, a brief cost-of-being-wrong line could be included to prevent overconfidence. 

\subsection{Limitations and Interpretation of Our Work} \label{limitations}

Our work has four main limitations. First, we inferred mental processes by administering psychometric questionnaires. We do not actually know what is happening inside people's heads. Think-aloud protocols in laboratory settings may be needed for that. Moreover, our results are contingent on the context of financial investments. Extending our model to other high-stakes domains (e.g., health) may test its generalizability. Additionally, the trust scales were administered at the conclusion of the task. However, our measures may potentially reflect not only the experimental interaction but also participants' prior experience with AI systems. As a result, we cannot attribute trust levels exclusively to the experimental manipulation. Nevertheless, any carryover from prior experience should be random with respect to our randomized conditions, and thus unlikely to bias condition comparisons. Future work may include a brief pre-task measure of the level of trust in AI to partial out pre-existing attitudes. Finally, domain knowledge was self-reported (i.e., measured via a questionnaire) to indicate a \textit{perceived} level of financial knowledge. This self-reported measure of domain knowledge is common in behavioral research~\citep{chen2022informal,flynn1999short,brucks1985effects}. Moreover, it aligns with our perception-driven model: like perceived anthropomorphism, trust, and perceived risk, it captures the user's felt competence -- the construct that actually guides judgment and behavior. Nevertheless, we acknowledge that this self-assessed construct may also reflect adjacent processes (e.g., self-confidence or familiarity) rather than objective expertise. To triangulate the construct, future work could compare finance professionals with lay participants.  

In our work, perceived anthropomorphism and risk perception were all measured as self-reported psychological states within the same experimental session. Accordingly, our use of trust as an intermediate variable may not be interpreted as establishing a temporal or causal sequence among these constructs. Rather, we employ mediation-style analyses as an analytical decomposition to examine how different trust-related perceptions are statistically associated with variations in perceived risk under differing levels of experienced anthropomorphism. This approach is consistent with prior human-AI interaction research that models trust as a risk-relevant psychological state rather than a manipulable decision variable~\citep{LiWuHuangLuan2024TrustworthyAI,GliksonWoolley2020TrustAI}. From this perspective, cognitive and affective trust may not ``cause'' changes in risk perception, but they certainly represent complementary dimensions of users' evaluative stance toward the AI system that co-vary with how risky reliance on its advice is perceived to be. Our findings therefore speak to patterns of association and differentiation among these subjective constructs, rather than to a mechanistic account of trust formation or causal risk reduction. We emphasize that longitudinal or experimental manipulation of trust itself may be required to make stronger claims about causal ordering, which remains an important direction for future work.

Despite these limitations, the present work makes a substantive contribution by empirically characterizing how perceived anthropomorphism, differentiated trust dimensions, and subjective risk judgments systematically co-vary in a high-stakes AI decision-support context, thereby offering robust, perception-level evidence that informs risk calibration and responsible design even in the absence of strong causal claims.

\section{Conclusion}
This study offers an evidence-based account of how anthropomorphism in AI affects risk perception via two trust pathways (cognitive and affective), and how domain knowledge shapes these relationships. The findings indicate that design choices associated with human-likeness do not operate uniformly: cues that elevate warmth do not necessarily support competence, and users with greater self-assessed knowledge weight these signals differently. For practice, the implication is to prioritize cues that convey competence, clarity, and verifiability, while using warmth as a complementary element rather than the primary driver. For strategy, adapt presentation to user expertise: what is reassuring for novices may be unpersuasive for experienced users, and vice-versa. For research, the model we estimated is a testable starting point, encouraging follow-up experiments with alternative manipulations, objective knowledge measures, and deployment in other high-stakes domains. Progress in this space will come from transparent evidence displays, calibrated explanations, and user-contingent presentation policies, not from a uniform increase in human-like styling.

\appendix
\section{Distribution Visualizations Across Conditions}
\label{app1}

\begin{figure*}[t]
\centering
\includegraphics[width=0.7\linewidth]{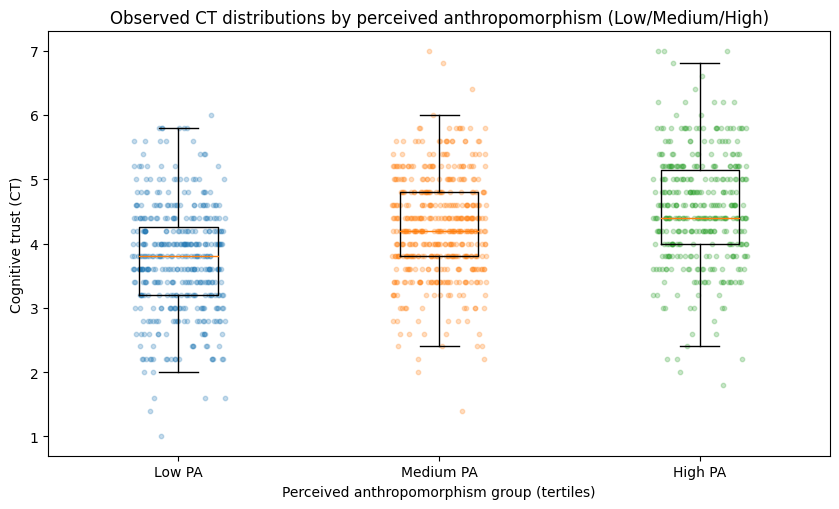}
\caption{Observed cognitive trust (CT) distributions across perceived anthropomorphism (PA) groups. PA was discretised into tertiles (Low, Medium, and High PA). Boxplots show the median and interquartile range of CT within each PA group (whiskers indicate dispersion), and semi-transparent jittered points represent individual participant observations. This figure provides an observed-data view of how CT varies across levels of perceived anthropomorphism, including within-group variability.}
\label{fig:dist1}
\end{figure*}

Figure~\ref{fig:dist1} visualises the raw distribution of cognitive trust (CT) across three perceived anthropomorphism (PA) groups (tertiles) to make the underlying data and dispersion explicit. The boxplots and overlaid individual observations show that CT tends to be higher when participants perceive the agent as more anthropomorphic (moving from low to high PA), while also revealing substantial within-group variability. This pattern aligns with the inferential results in which PA positively predicts CT, and it clarifies that the relationship reflects a shift in the central tendency of CT rather than being driven by a small number of extreme observations.

\begin{figure*}[t]
\centering
\includegraphics[width=0.7\linewidth]{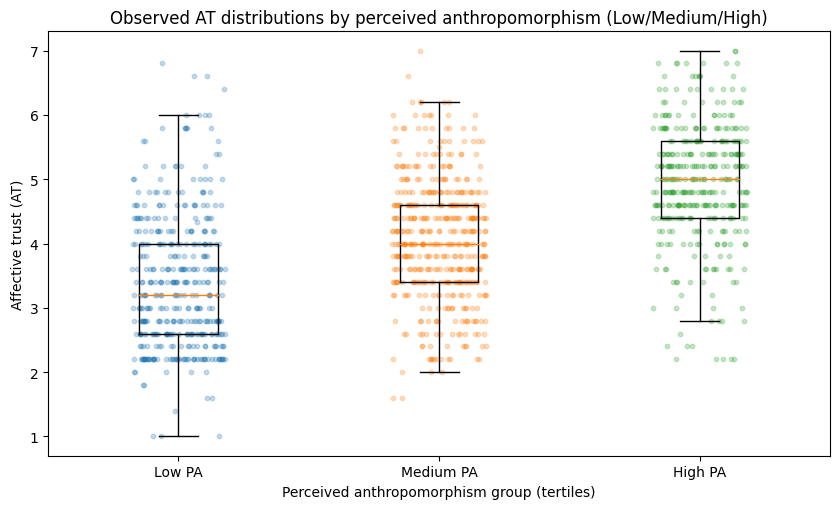}
\caption{Observed affective trust (AT) distributions across perceived anthropomorphism (PA) groups. PA was discretised into tertiles (Low/Medium/High PA). Boxplots summarise the distribution of AT within each PA group, and semi-transparent jittered points show individual observations. This figure illustrates the observed association between perceived anthropomorphism and affective trust, including the extent of variability within each PA level.}
\label{fig:dist2}
\end{figure*}

Figure~\ref{fig:dist2} presents the raw distribution of affective trust (AT) across low, medium, and high PA tertiles, and it is used to show observed data rather than only model-predicted slopes. The distributional shift across PA groups indicates that participants who perceive the agent as more anthropomorphic also tend to report higher affective trust, consistent with the hypothesised effect of PA on AT. At the same time, the visible spread of points within each group highlights meaningful heterogeneity in affective responses, suggesting that anthropomorphism perceptions increase AT on average while leaving room for individual differences in how social cues translate into relational trust.

\section{Robustness Checks for Interaction Analyses}
\label{app2}
The analysis below presents additional testing of the interaction between anthropomorphism and the two trust-based risks identified in the regression model. Along with the main interaction terms PA $\times$ DK present in the main model, we tested two interaction terms on trust -- i.e., we extended the final risk-perception model (Eq.~1 / Model~5; controls + PA + CT + AT + DK + PA$\times$DK) by adding two additional interaction terms, PA$\times$CT and PA$\times$AT. The extended model was estimated using OLS with robust (HC3) standard errors. All continuous predictors were mean-centered prior to computing product terms.

\paragraph{Model specification}


\begin{equation}
\label{eq:rp_step4}
\begin{split}
\text{RP} \sim\;& \text{Controls} + \text{PA} + \text{CT} + \text{AT} + \text{DK} \\
&+ (\text{PA}\times \text{DK}) + (\text{PA}\times \text{CT}) \\
&+ (\text{PA}\times \text{AT}).
\end{split}
\end{equation}

\paragraph{Model fit}
Adding PA$\times$CT and PA$\times$AT produced a small improvement in model fit: the base model (Eq.~1 + PA$\times$DK) explained $R^2 = 0.0731$ of the variance in RP, whereas the extended model explained $R^2 = 0.0782$ ($\Delta R^2 = 0.0051$). The AIC also decreased, indicating a better-fitting specification.

\paragraph{Key coefficients}
In the extended model, PA remained a positive predictor of RP ($b=0.096$, $p=.0109$), while CT remained a strong negative predictor ($b=-0.342$, $p<.001$). AT was not a significant predictor of RP ($b=0.078$, $p=.148$), whereas DK was positively associated with RP ($b=0.114$, $p<.001$). The PA$\times$DK interaction was not significant ($b=0.024$, $p=.202$). The two reviewer-requested interactions were marginal: PA$\times$CT was negative ($b=-0.073$, $p=.062$) and PA$\times$AT was positive ($b=0.059$, $p=.060$).

\paragraph{Interpretation}
Although these interaction effects did not reach conventional significance levels, they exhibit a theoretically coherent pattern. The negative PA$\times$CT coefficient suggests that higher cognitive trust attenuates the association between perceived anthropomorphism and risk perception (i.e., when the system is viewed as more competent/reliable, anthropomorphic perceptions contribute less to risk judgments). Conversely, the positive PA$\times$AT coefficient suggests that higher affective trust may amplify the PA--RP association (i.e., relational trust may increase sensitivity to anthropomorphic impressions when evaluating risk in a high-stakes context). Importantly, including these interactions did not alter the core inferences of the primary mediation model.

\paragraph{Multicollinearity}
After mean-centering, collinearity remained low (maximum VIF $\approx 1.82$), indicating that the interaction tests are unlikely to be artifacts of multicollinearity.

\begin{figure*}[t]
\centering
\includegraphics[width=0.7\linewidth]{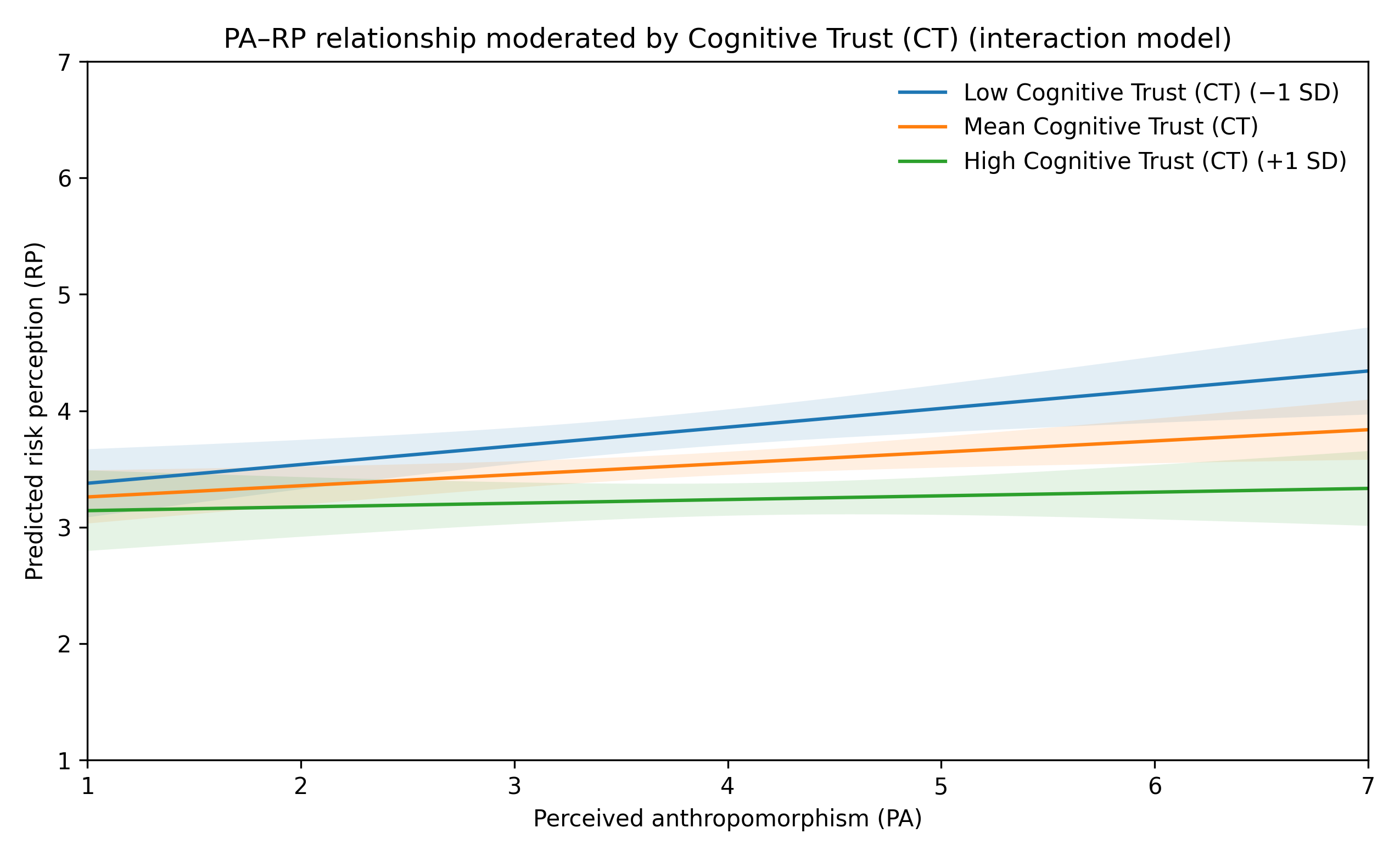}
\caption{Interaction plot for PA $\times$ Cognitive Trust (CT). Predicted risk perception (RP) as a function of perceived anthropomorphism (PA) at low ($-1$ SD), mean, and high ($+1$ SD) levels of cognitive trust (CT), based on the extended regression model. Shaded regions denote 95\% confidence intervals. The PA--RP slope is strongest when CT is low and becomes flatter as CT increases, consistent with a negative PA$\times$CT interaction (attenuation of the PA--RP association at higher cognitive trust).}
\label{fig:intCT}
\end{figure*}

\begin{figure*}[t]
\centering
\includegraphics[width=0.7\linewidth]{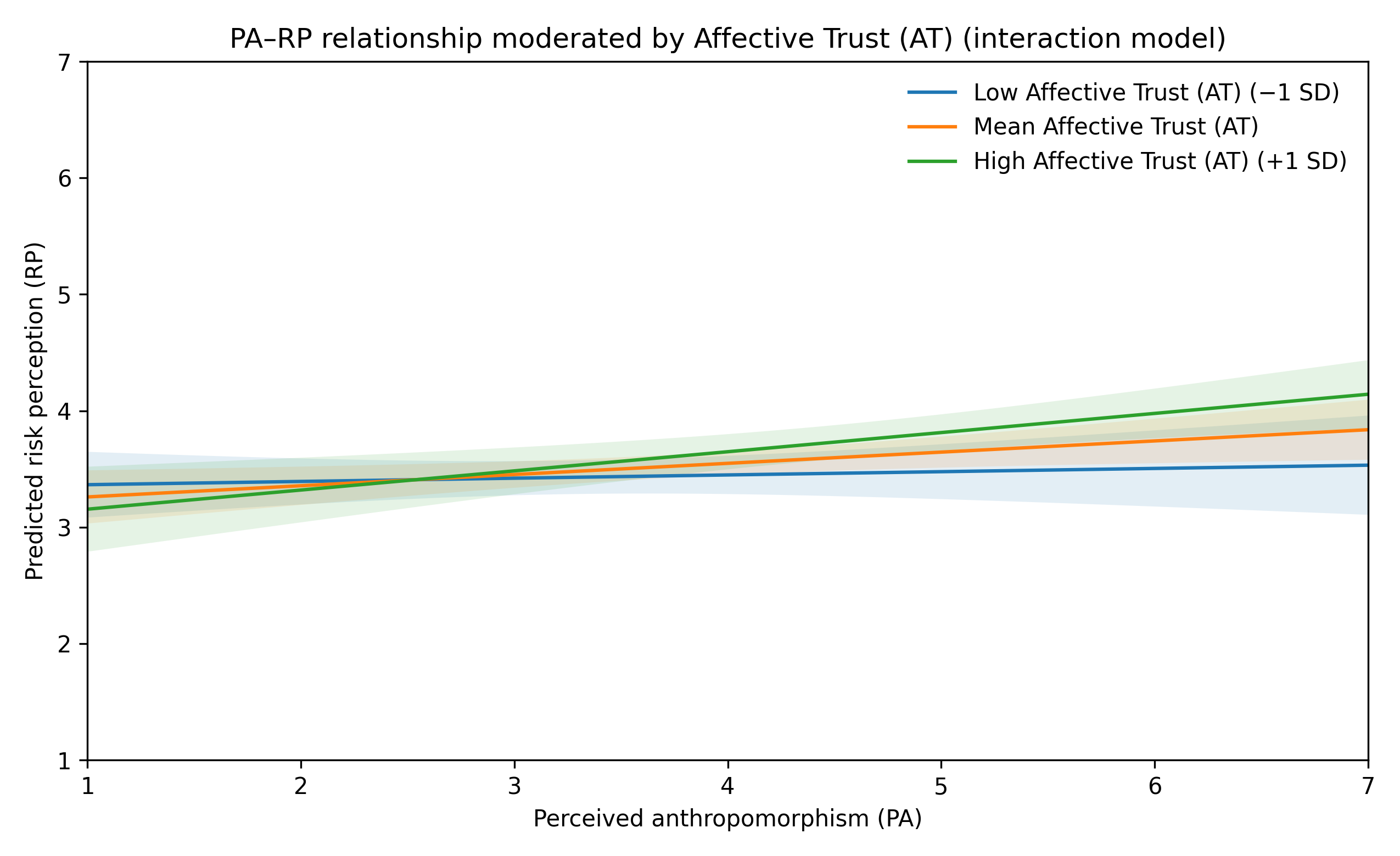}
\caption{Interaction plot for PA $\times$ Affective Trust (AT). Predicted risk perception (RP) as a function of perceived anthropomorphism (PA) at low ($-1$ SD), mean, and high ($+1$ SD) levels of affective trust (AT), based on the extended regression model. Shaded regions denote 95\% confidence intervals. The PA--RP slope becomes steeper as AT increases, consistent with a positive PA$\times$AT interaction (amplification of the PA--RP association at higher affective trust).}
\label{fig:intAT}
\end{figure*}

\paragraph{Interaction Plots}
Figure~\ref{fig:intCT} and Figure~\ref{fig:intAT} visualise the interaction tests. Each plot shows the predicted relationship between perceived anthropomorphism (PA) and risk perception (RP) at three levels of the trust moderator (low = $-1$ SD, mean, high = $+1$ SD), holding all other predictors at their sample means. Shaded bands represent 95\% confidence intervals. These visualisations indicate that the PA--RP association varies with trust: the slope becomes flatter as cognitive trust increases (consistent with the negative PA$\times$CT coefficient), whereas the slope becomes steeper as affective trust increases (consistent with the positive PA$\times$AT coefficient). While the corresponding interaction terms are marginal, the plots provide an intuitive representation of their direction and magnitude.

\bibliographystyle{elsarticle-harv} 

\bibliography{reference}





\end{document}